\documentclass[preprint,12pt]{elsarticle}

\usepackage{graphicx}
\usepackage{amssymb}
\usepackage{url}
\usepackage{subfig}
\usepackage{color}
\usepackage{hyperref}
\usepackage{units}
\usepackage{soul}
\usepackage[top=1in, bottom=1in, left = 1in, right = 1in]{geometry}

\biboptions{square, sort&compress}

\journal{Computational Materials Science}

\begin{document}

\begin{frontmatter}


\title{Benchmark Problems for Numerical Implementations of Phase Field Models}

\author[CHiMaD]{A. M. Jokisaari}
\author[CHiMaD,NU]{P. W. Voorhees}
\author[NIST]{J. E. Guyer}
\author[NIST]{J. Warren}
\author[NAISE,ANL]{O. G. Heinonen}
\address[CHiMaD]{Center for Hierarchical Materials Design, Northwestern University, 2205 Tech Drive, Evanston, IL, 60208}
\address[NU]{Department of Materials Science and Engineering, Northwestern University, 2220 Campus Drive, Evanston, IL 60208}
\address[NIST]{Material Measurement Laboratory, National Institute of Standards and Technology, 100 Bureau Drive, MS 8300, Gaithersburg, MD 20899-8300}
\address[NAISE]{Northwestern-Argonne Institute of Science and Engineering, Evanston, Illinois 60208, USA}
\address[ANL]{Materials Science Division, Argonne National Laboratory, Lemont, IL 60439}

\begin{abstract}

We present the first set of benchmark problems for phase field models that are being developed by the Center for Hierarchical Materials Design (CHiMaD) and the National Institute of Standards and Technology (NIST).  While many scientific research areas use a limited set of well-established software, the growing phase field community continues to develop a wide variety of codes and lacks benchmark problems to consistently evaluate the numerical performance of new implementations. Phase field modeling has become significantly more popular as computational power has increased and is now becoming mainstream, driving the need for benchmark problems to validate and verify new implementations. We follow the example set by the micromagnetics community to develop an evolving set of benchmark problems that test the usability, computational resources, numerical capabilities and physical scope of phase field simulation codes. In this paper, we propose two benchmark problems that cover the physics of solute diffusion and growth and coarsening of a second phase via a simple spinodal decomposition model and a more complex Ostwald   ripening model. We demonstrate the utility of benchmark problems by comparing the results of simulations performed with two different adaptive time stepping techniques, and we discuss the needs of future benchmark problems. The development of benchmark problems will enable the results of quantitative phase field models to be confidently incorporated into integrated computational materials science and engineering (ICME), an important goal of the Materials Genome Initiative.
\end{abstract}

\begin{keyword}
Phase field model \sep Benchmark problem \sep Spinodal decomposition \sep Ostwald ripening
\end{keyword}

\end{frontmatter}

%\linenumbers

\section{Introduction}
\label{S:Introduction}

Many important processes in materials microstructural evolution, such as coarsening, solidification, polycrystalline grain evolution, and magnetic and ferroelectric domain formation and motion, occur on mesoscopic length and time scales. The ``mesoscale'' is the scale ``in between;" in this case, in between atomistic scales of the order of sub-nanometers and femto- to picoseconds, and macroscopic scales of the order of micrometers and microseconds and larger. Mesoscale processes can strongly impact materials properties and performance in engineering applications, providing strong motivation to develop accurate mesoscale microstructure evolution models.

Two general mesoscale modeling approaches exist, with the primary difference being how interfaces are handled \cite{moelans2008introduction,emmerich2008advances,duddu2011diffusional}.  Sharp-interface approaches, which treat interfaces as mathematically sharp, can be very efficient numerically when simulating the evolution of simple microstructural geometries. However, interface tracking with complex geometries (e.g., during dendritic growth) and topology changes, such as particles merging or splitting, pose significant numerical challenges \cite{duddu2011diffusional}.  Diffuse-interface approaches, in which the interface has a finite width, avoid these issues \cite{moelans2008introduction,emmerich2008advances,duddu2011diffusional}. However, they generally require more computational resources because the diffuse interface, which often has a width of a few nanometers, must be resolved even as other structural features may have length scales in the hundreds of nanometers or larger. 

One popular diffuse-interface technique is the phase field approach, which has been used to study dendritic growth, spinodal decomposition, grain growth, ferroelectric domain formation, and other phenomena \cite{ASM,boettinger2002phase,chen2002phase,emmerich2008advances,moelans2008introduction,steinbach2009phase,nestler2011phase,steinbach2013phase}. In a phase field model, a microstructure is described by one or more continuous fields, $\varphi\left(\mathbf{r},t\right)$.  The fields change smoothly over the computational domain and across interfaces. The field variables may be either a physical quantity, such as composition or density, or a phenomenological descriptor \cite{moelans2008introduction}. Originally \cite{Fix,Langer}, the fields were used to denote a local phase (hence the name {\em phase field}), with the value of $\varphi$ at position $\mathbf{r}$ and time $t$ indicating the phase. For example, a two-phase system can be described by a field $\varphi$ that takes the values $\varphi_{\alpha}$ and $\varphi_{\beta}$ in the bulk $\alpha$ and $\beta$ phases, respectively, while at the $\alpha$/$\beta$ interface, the value of $\varphi$ changes smoothly over a finite width. The use of phase field methods is now more diverse, with the phase field variable often representing other quantities or properties, such as concentration or density. The evolution of existing phases within the system is driven by the reduction of the free energy, which is described as a functional of the field variables.  Depending on the physics being modeled, the field variables may be conserved or non-conserved. Finally, ``sharp-interface limit'' or ``thin-interface limit'' analyses have shown that phase field models are equivalent to their analogous sharp-interface models when the interface width is significantly smaller than the size of other characteristic length scales (reviewed in Refs. \cite{moelans2008introduction,emmerich2008advances}). For comprehensive descriptions and reviews of phase field modeling, see Refs.\ \cite{ASM,boettinger2002phase,chen2002phase,emmerich2008advances,moelans2008introduction,steinbach2009phase,nestler2011phase,steinbach2013phase}.

Quantitative phase field models have been developed to study technologically important phenomena in real materials systems as part of integrated computational materials engineering (ICME) \cite{furrer2011application,luo2015material,schmitz2015microstructure}. In ICME, models at different length scales are linked together to design materials for technological applications. A few selected references of recent quantitative phase field studies include solidification in Al alloys \cite{qin2003phase,kobayashi2003phase,bottger2009simulation}, precipitation in Ni-based superalloys \cite{zhu2002linking,zhu2004three,kitashima2009new}, recrystallization in Ti \cite{gentry2015simulating} and Mg \cite{wang2009grain} alloys,  quantum dot formation in  InAs/GaAs \cite{aagesen2012phase}, and semiconducting core-shell nanoparticles \cite{mangeri2015influence}.  The phase field approach continues to be applied to novel materials systems and phenomena, and a growing number of scientists are adopting the technique. 

The number of phase field software implementations is proliferating with the growing application of phase field techniques, necessitating a means of benchmarking, validating, and verifying the numerical behavior of a diverse set of codes. Many research domains which apply computational modeling have converged around a small number of standard pieces of software and benchmarking sets (e.g., COMSOL \cite{comsol2015comsol} and ABAQUS \cite{abaqus} for engineering simulations, or VASP \cite{kresse1993ab,kresse1994ab,kresse1996efficiency,kresse1996efficient}, Quantum ESPRESSO \cite{quantumEspresso}, and the 
G3/99 test set \cite{Curtiss_JCP2000} for electronic structure calculations\footnote{Certain commercial equipment, instruments, or materials are identified in this paper to foster understanding. Such identification does not imply recommendation or endorsement by the National Institute of Standards and Technology, nor does it imply that the materials or equipment identified are necessarily the best available for the purpose.}), but this is not the case for the phase field community. A multitude of phase field software implementations exist, and numerical approaches abound. Phase field simulations have been performed using open-source codes such as MOOSE \cite{tonks2012object, millett2013three}, FEniCS \cite{fenics,  welland2015miscibility}, OpenPhase \cite{steinbach2009phase}, DUNE \cite{bastian2008generic,bastian2008genericII}, FiPy \cite{fipy,Wheeler:2010p2378}, as well as with many proprietary codes, such as MICRESS \cite{steinbach1996phase,mecozzi2016phase}, PACE 3D \cite{nestler2005multicomponent, stinner2004diffuse} and other in-house codes.  Numerical implementations may employ finite difference, finite volume, finite element, or spectral methods to solve the evolution equations, direct or spectral methods for solid mechanics calculations, explicit or implicit time stepping, and adaptive or non-adaptive meshing. To confidently incorporate quantitative phase field results obtained from this wide variety of numerical methods into ICME, both physical models and numerical implementations must be validated and verified.   

A set of standard benchmark problems allows the comparison of models, algorithms, and implementations, as well as the testing of solution accuracy, solver optimizations, and code modifications. While the phase field community ultimately needs validated experimental data sets to compare different models, we focus our effort here on first developing benchmark problems for numerical implementations, which is a necessary precursor for the comparison of model results; a model cannot be validated in a useful way until questions about the correctness of numerical implementations are resolved. The micromagnetics community created benchmark problems in the late 1990s to early 2000s to address a similar situation of multiple implementations and numerical methods  \cite{muMAG}, and these problems are still evolving today. Benchmark problems significantly aided the community in creating accurate micromagnetics codes \cite{muMAG}, such as the Object Oriented MicroMagnetics Framework (OOMMF) \cite{donahue1999oommf}, MuMax3 \cite{vansteenkiste2014design}, and Magpar \cite{scholz2003scalable}. To aid in the development, validation, and verification of phase field modeling software, the Center for Hierarchical Design (CHiMaD) and the National Institute of Standards and Technology (NIST) are developing phase field benchmark problems. These problems are hosted on the NIST website \cite{PFBenchmark} and are freely available. In addition, NIST will also host the solutions to the problems submitted by members of the phase field community so that the results from different implementations may be compared. 

Phase field benchmark problems for numerical implementations should exhibit several key features, analogous to those in the micromagnetics benchmark problems. First, the problems should be nontrivial (i.e., not solvable without a computer) and should exhibit differing degrees of computational complexity, yet not require extensive computational resources. Second, simulation outputs must be defined in such a way that results are easily comparable. In addition to snapshots or videos of the evolution of the microstructure itself, the evolution of overall metrics such as the total energy of the system or the volume fraction of each phase should be quantified. Finally, the problems should test a simple, targeted aspect of either the numerical implementation or the physics. For example, simple physics could be used while complicated domain or boundary conditions are tested, or coupled physics could be tested on a simple domain. Numerical aspects that must be challenged include solver algorithms, mesh geometry, boundary conditions, and time integration. Benchmark problems could be especially useful when examining multiphysics coupling, including such behaviors as, e.g.,  diffusion, linear elasticity, fluid flow, anisotropic interfacial energy, and polarization.

In this paper, we present a first set of community-driven, benchmark problems for numerical implementations of phase field models and the efforts of NIST and CHiMaD to date. This first set of problems focuses on diffusion of a solute and phase separation; the second problem adds a coupled non-conserved order parameter. We discuss our choice of model formulations, parameterizations and initial conditions so that these considerations may be kept in mind while developing additional benchmark problems.  Furthermore, we demonstrate the utility of benchmark problems by comparing simulation results obtained using two different time adaptivity algorithms.  We also briefly review lessons learned from the first CHiMaD ``Hackathon,'' an event in which different phase field codes within the community were challenged against model problems. Finally, we discuss the development of additional formulations for the future, and encourage community involvement in the entire process of problem design, development, and reporting of results.

\section{Model formulations \label{sec:Model-formulations}}

In phase field models, field variables are evolved using dynamics derived from generalized forces. The field variable is often termed the ``order parameter,'' and we adopt that terminology here. Most commonly, the time evolution is governed by dissipative dynamics, in which the total free energy of the system decreases monotonically with time (i.e., entropy increases at fixed temperature).  The order parameter may be locally conserved or non-conserved depending on what physical quantity or property the order parameter represents, and its dynamics are defined by the response of the system to a generalized force defined by the variation in the free energy. Kinetic coefficients, such as mobility or diffusivity, control how the order parameter responds to the force. An example of a conserved order parameter is the concentration of solute in a matrix, while ferroelectric polarization is an example of a non-conserved order parameter. 

The first problem in this benchmark set models spinodal decomposition via conserved dynamics, while the second models Ostwald ripening via coupled conserved/non-conserved dynamics. In this way, we focus on a single, fundamental aspect of physics (i.e., diffusion and phase separation) in the first problem, and then increase the model complexity in the second problem. We discuss the motivation for each model formulation and the choice of initial conditions, boundary conditions, and computational domains. The problems were formulated to be effectively two-dimensional so that the essential physical behavior is modeled without making the test problems unreasonably large or computationally demanding. 

\subsection{Spinodal decomposition \label{sub:Spinodal-decomposition}}

Spinodal decomposition is one of the oldest problems in the phase field canon, and its formulation in terms of continuum fields goes back to the seminal works by Cahn and Hilliard \cite{cahn1961spinodal}. The Cahn-Hilliard equation thus predates the name ``phase field'' in this context, but the term has subsequently  been adopted by the community. While spinodal decomposition may be one of the simplest problems to model, it is highly relevant, as a large number of phase field models include the diffusion of a solute within a matrix. Furthermore, precipitation and growth may also be modeled with the same formulation if the appropriate initial conditions are chosen. For the benchmark problem, we select a simple formulation that is numerically tractable so that results may be obtained quickly and interpreted easily, testing the essential physics while minimizing model complexity and the chance to introduce coding errors.

\subsubsection{Free energy and dynamics}\label{sssec:spinodal_free_energy}

For this benchmark problem of spinodal decomposition in a binary system, a single order parameter, $c$, is evolved, which describes the atomic fraction of solute. The free energy of the system, $F$, is expressed as \cite{cahn1961spinodal}
\begin{equation}
F=\int_{V}\left(f_{chem}\left(c\right)+\frac{\kappa}{2}|\nabla c|^{2}\right)dV,\label{eq:F_spinodal}
\end{equation}
where $f_{chem}$ is the chemical free energy density and $\kappa$ is the gradient energy coefficient. For this problem, we choose $f_{chem}$ to have a simple polynomial form,
\begin{equation}
f_{chem}\left(c\right)=\varrho_{s}\left(c-c_{\alpha}\right)^{2}\left(c_{\beta}-c\right)^{2},\label{eq:fchem_spinodal}
\end{equation}
such that $f_{chem}$ is a symmetric double-well with minima at $c_{\alpha}$ and $c_{\beta}$, and $\varrho_{s}$ controls the height of the double-well barrier. Because $f_{chem}$ is symmetric (Fig.\ \ref{fig:energy_p1}), $c_{\alpha}$ and $c_{\beta}$ correspond exactly with the equilibrium atomic fractions of the $\alpha$ and $\beta$ phases. 

Because $c$ must obey a continuity equation -- the flux of $c$ is conserved -- the evolution of $c$ is given by the Cahn-Hilliard equation \cite{cahn1961spinodal}, which is derived from an Onsager force-flux relationship \cite{balluffi2005kinetics}:

\begin{equation}
\frac{\partial c}{\partial t}=\nabla\cdot\Bigg\{M\nabla\left(\frac{\partial f_{chem}}{\partial c}-\kappa\nabla^{2}c\right)\Bigg\} \label{eq_fullCH_p1}
\end{equation}
where $M$ is the mobility of the solute. For simplicity, both the mobility and the interfacial energy are isotropic. We choose $c_{\alpha}=0.3$, $c_{\beta}=0.7$, $\varrho_{s}=5$, $M=5$, and $\kappa=2$. Because the interfacial energy, diffuse interface width, and free energy parameterization are coupled, we obtain the diffuse interface width of $l=7.071 \sqrt{\kappa/\varrho_s}=4.47$ units over which $c$ varies as  $0.348<c<0.652$, and an interfacial energy $\sigma=0.01508\sqrt{\kappa \varrho_s}$ \cite{cahn1958free}.

\begin{center}
\begin{figure}
\begin{centering}
\subfloat[\label{fig:energy_p1}]{\begin{centering}
\includegraphics[scale=0.95]{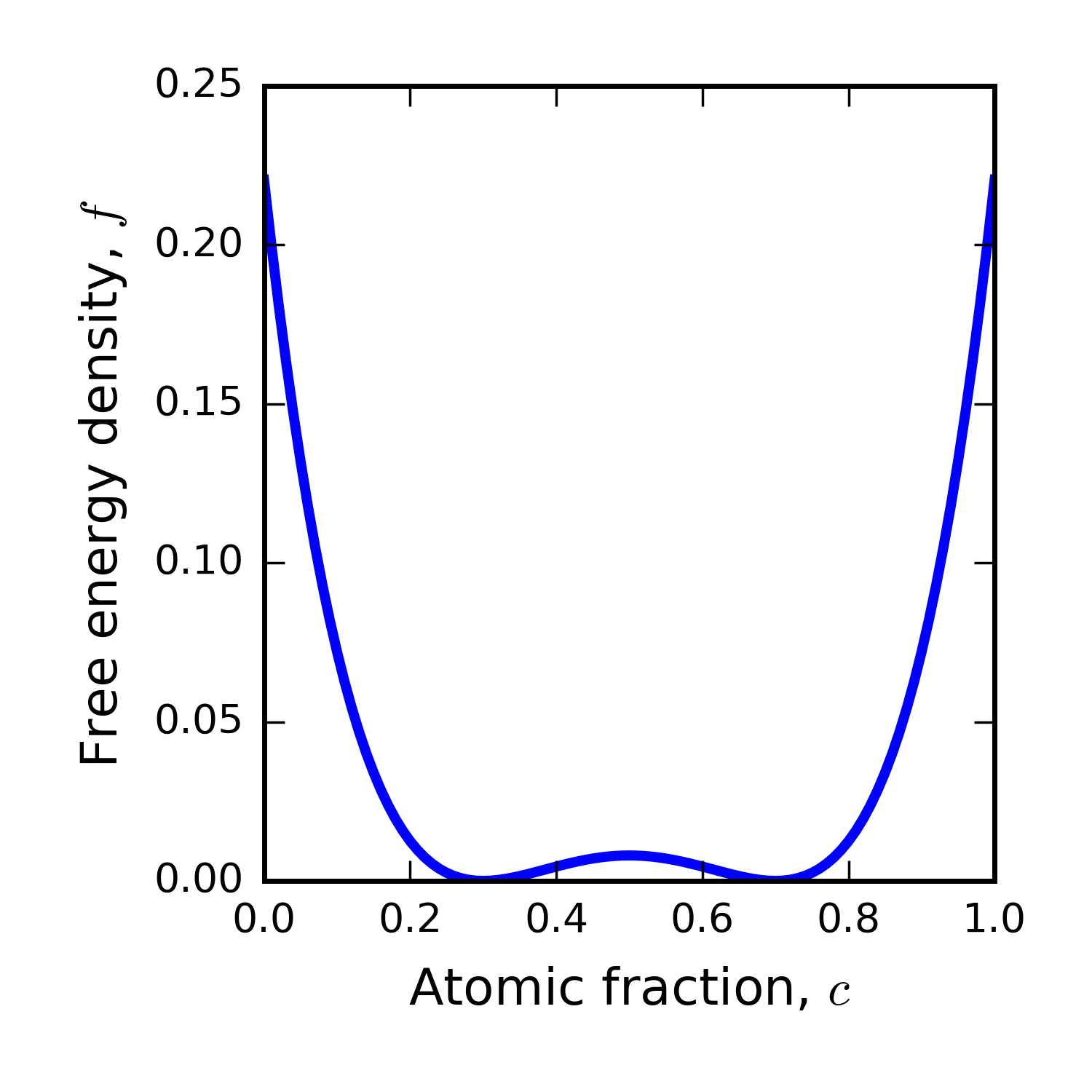}
\par\end{centering} 
}
\subfloat[\label{fig:energy_p2}]{\begin{centering}
\includegraphics[scale=0.65]{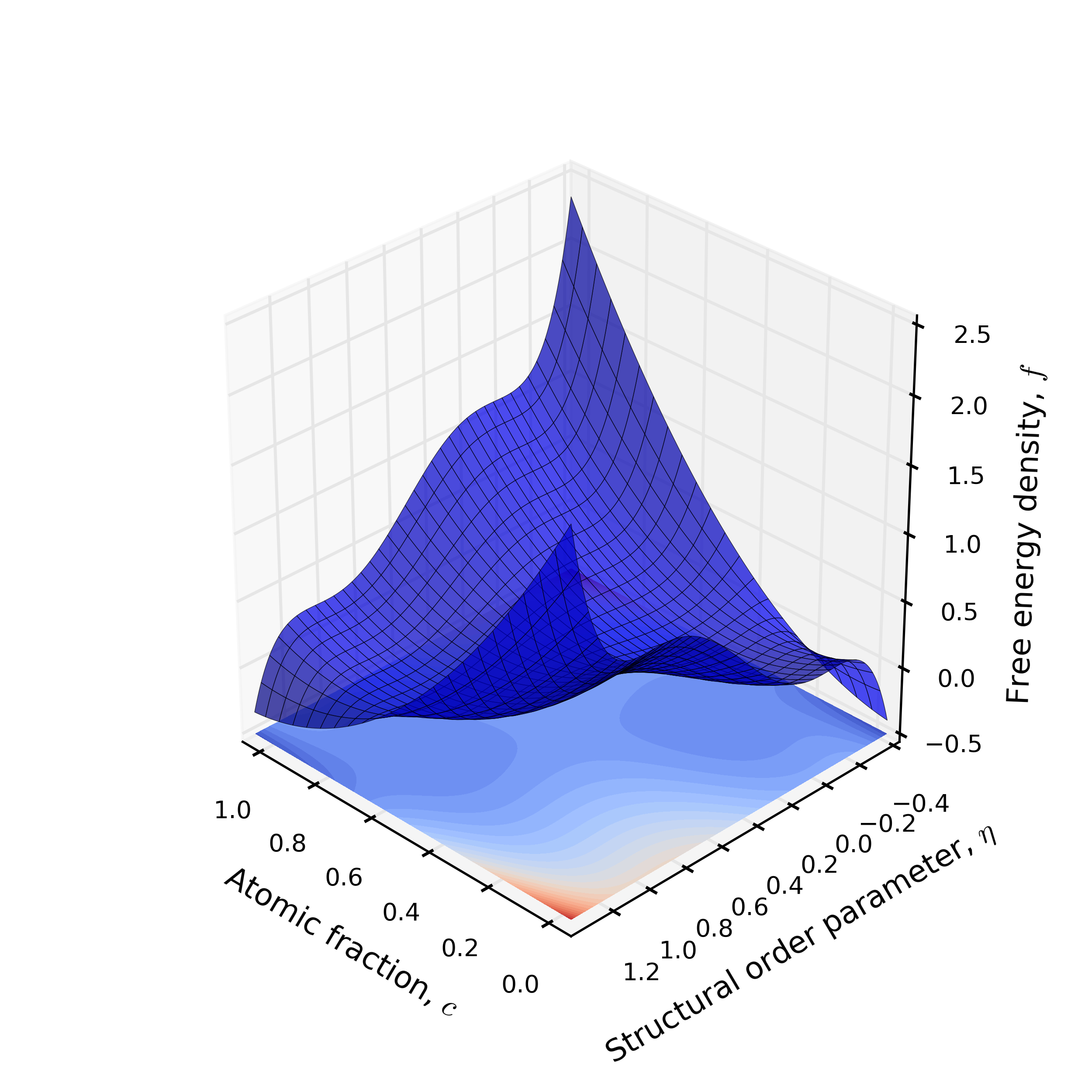}
\par\end{centering}
}

\par\end{centering}

\caption{The free energy density surfaces for a) the spinodal decomposition problem, and b) the Ostwald ripening problem for $(c,\ \eta)$ (not shown: $(\eta_i, \eta_j)$ surface). The free energy density surfaces are defined for all real values of $c$ (both problems) and $\eta_i$ (Ostwald ripening problem) and not only over the intervals of interest, necessitating care in choosing initial conditions (Section \ref{sub:choices}). \label{fig:free energy surfaces}}
\end{figure}
\par\end{center}

\subsection{Ostwald ripening \label{sub:Ostwald-ripening}}

The second benchmark problem examines Ostwald ripening in a system with an ordered phase and a disordered phase; an example of this phenomenon in a real materials system is the growth and coarsening of $\gamma'$
precipitates in a $\gamma$ matrix in nickel-based superalloys \cite{pollock2006nickel,zhu2002linking}. This system is somewhat more complicated than that presented in Section \ref{sub:Spinodal-decomposition}, in that the microstructural evolution is driven by coupled conserved/non-conserved dynamics. However, the
formulation is a simple extension of that in the previous section (note that we neglect elastic energy, an important factor in $\gamma$/$\gamma'$ evolution).

\subsubsection{Free energy and dynamics}\label{sssec:ostwald_free_energy}

The atomic fraction of solute is again specified by the conserved variable $c$, while the phase is indicated by a structural order parameter, $\eta$.  The structural order parameter is non-conserved and is a phenomenological phase descriptor, such that the $\alpha$ phase is indicated by $\eta=0$, while the $\beta$ phase is indicated by $\eta=1$. If multiple energetically equivalent orientation variants exist (for example, due to crystallographic symmetry considerations or ordered and disordered phases), the model may include $p$ number of structural order parameters, $\eta_{p}$, with one for each orientation variant.  We include a nontrivial number of order parameters by setting $p=4$, a value commonly used in superalloy models; this will stress the numerical solver while not making the problem intractable.

In this benchmark problem, the free energy of the system is based on the formulation presented in Ref.\ \cite{zhu2004three} and is expressed as 
\begin{equation}
F=\int_{V}\left(f_{chem}\left(c,\eta_{1},...\eta_{p}\right)+\frac{\kappa_{c}}{2}|\nabla c|^{2}+\sum_{i=1}^{p}\frac{\kappa_{\eta}}{2}|\nabla\eta_{i}|^{2}\right)dV\label{eq:p2_F}
\end{equation}
where $\kappa_{c}$ and $\kappa_{\eta}$ are the gradient energy coefficients for $c$ and $\eta_{i}$, respectively. While the model in Ref.\ \cite{zhu2004three} follows the Kim-Kim-Suzuki (KKS) formulation for interfacial energy \cite{kim1999phase}, we use the Wheeler-Boettinger-McFadden (WBM) \cite{wheeler1992phase}
formulation for simplicity.  In the KKS model, the interface is treated as an equilibrium mixture of two phases with fixed compositions such that an arbitrary diffuse interface width may be specified for a given interfacial energy.  In the WBM model,  interfacial energy and interfacial width are linked with the concentration, such that very high resolution across the interface may be required to incorporate accurate interfacial energies. 

The formulation for $f_{chem}$ in Ref.\ \cite{zhu2004three} is adapted for our benchmark problem as
\begin{equation}
f_{chem}\left(c,\eta_{1},...\eta_{p}\right)=f^{\alpha}\left(c\right)\left[1-h\left(\eta_{1},...\eta_{p}\right)\right]+f^{\beta}\left(c\right)h\left(\eta_{1},...\eta_{p}\right)+wg\left(\eta_{1},...\eta_{p}\right),\label{eq:p2_fchem}
\end{equation}
where $f^{\alpha}$ and $f^{\beta}$ are the chemical free energy densities of the $\alpha$ and $\beta$ phases, respectively, $h\left(\eta_{1},...\eta_{p}\right)$ is an interpolation function, and $g\left(\eta_{1},...\eta_{p}\right)$ is a double-well function. The function $h$ increases monotonically
between $h(0)=0$ and $h(1)=1$, while the function $g$ has minima at $g(0)=0$ and $g(1)=0$. The height of the double well barrier is controlled by $w$. We choose the simple formulation 
\begin{equation}
f^{\alpha}\left(c\right)=\varrho^{2}\left(c-c_{\alpha}\right)^{2}\label{eq:f_alpha}
\end{equation}
\begin{equation}
f^{\beta}\left(c\right)=\varrho^{2}\left(c_{\beta}-c\right)^{2}\label{eq:f_beta}
\end{equation}
\begin{equation}
h\left(\eta_{1},...\eta_{p}\right)=\sum_{i=1}^{p}\eta_{i}^{3}\left(6\eta_{i}^{2}-15\eta_{i}+10\right)\label{eq:h}
\end{equation}
\begin{equation}
g\left(\eta_{1},...\eta_{p}\right)=\sum_{i=1}^{p}\left[\eta_{i}^{2}\left(1-\eta_{i}\right)^{2}\right]+\alpha\sum_{i=1}^{p}\sum_{j\neq i}^{p}\eta_{i}^{2}\eta_{j}^{2},\label{eq:g}
\end{equation}
where $f^{\alpha}$ and $f^{\beta}$ have minima at $c_{\alpha}$
and $c_{\beta}$, $\varrho^{2}$ controls the curvature of the free
energies, and $\alpha$ controls the energy penalty incurred by the overlap of multiple non-zero $\eta_{i}$ values at the same point.  Because the energy values of the minima are the same (Fig.\ \ref{fig:energy_p2}), $c_{\alpha}$ and $c_{\beta}$ correspond exactly with the equilibrium atomic fractions of the $\alpha$ and $\beta$ phases. 

The time evolution of $c$ is again governed by the Cahn-Hilliard equation \cite{cahn1961spinodal,elliott1989second}, 

\begin{equation}
\frac{\partial c}{\partial t}=\nabla\cdot\Bigg\{M\nabla\left(\frac{\partial f_{chem}}{\partial c}-\kappa_{c}\nabla^{2}c\right)\Bigg\}. \label{eq:full_CH_p2}
\end{equation}
The Allen-Cahn equation \cite{allen1979microscopic}, which is based on gradient flow, governs the evolution of $\eta_{i}$, 

\begin{equation}
\frac{\partial\eta_{i}}{\partial t}=-L\left[\frac{\delta F}{\delta \eta_{i}}\right]=-L\left(\frac{\partial f_{chem}}{\partial\eta_{i}}-\kappa_{\eta}\nabla^{2}\eta_{i}\right),\label{eq:p2_AC}
\end{equation}
where $L$ is the kinetic coefficient of $\eta_{i}$. We choose $M=5$ and $L=5$ so that the transformation is diffusion-controlled, and as in Section \ref{sub:Spinodal-decomposition}, the kinetic coefficients
and gradient energy coefficients are isotropic. In addition, we again choose $c_{\alpha}=0.3$ and $c_{\beta}=0.7$, and further specify $k_{c}=k_{\eta}=3$, $\varrho=\sqrt[]{2}$, $w=1$, and $\alpha=5$.
For these values, the diffuse interface between $0.1<\eta<0.9$ has a width of of 4.2 units. 

\subsection{Reasons for choices of models and parameters \label{sub:choices}}

The two benchmark problems presented here are simplified formulations designed to focus on fundamental aspects common to almost every phase field model: the diffusion of solute (Section \ref{sub:Spinodal-decomposition}) and the coupling of composition with a structural order parameter (Section \ref{sub:Ostwald-ripening}). All of the model parameters chosen here are within a few orders of unity, improving numerical performance. In addition, the structural order parameter in the second model is phenomenological and varies within the interval of {[}0, 1{]}. This interval is chosen because multiphysics coupling that relies on the phase of the material is often incorporated by way of a structural order parameter (e.g., misfit strain of a precipitate phase with respect to a matrix phase). 

Several trade-offs were considered between the free energy formulations and the initial conditions. The free energy could be chosen to be realistic, for example by using the CALPHAD method, or to be more simplistic while still representing the main physics and being numerically tractable. The CALPHAD method is a semi-empirical method of formulating free energies of mixing using known thermodynamic data and equilibrium phase diagrams \cite{saunders1998calphad, lukas2007computational}. While CALPHAD free energies are extremely useful, their functional form generally contains natural logarithms, which pose several numerical and mathematical challenges for incorporation into phase field models. Therefore, simple polynomial free energy density formulations are chosen because they are numerically tractable and straightforward to implement. Ideally, the energy formulation should be robust such that the system will tend to the equilibrium values of the phases no matter the initial condition.  This behavior may be ensured by fixing the global minimum of the free energy density within the interval of interest. However, many formulations do not exhibit these global minima, including the one in Section \ref{sub:Ostwald-ripening} (Fig.\ \ref{fig:free energy surfaces}b). In addition, the characteristics of the free energy density surface are sensitive to the parameterization of the model. For example, the local minima present at $\eta=0$, $c=0.3$ and $\eta=1,$ $c=0.7$ become shallower as $w$ decreases. For certain values of $w$ and $\alpha$ (e.g., $w=0.1$ and $\alpha=1$), the lowest energy occurs when all of the structural order parameters assume a value of approximately 0.9 in the $\beta$ phase. This behavior is due to the $\eta_{i}^{2}\eta_{j}^{2}$ term in $g$. Finally, transient solute depletion in the $\alpha$ phase, which may cause $c$ to decrease below 0, may occur during the first several time steps of the simulation as the system quickly relaxes from its initial conditions.  Furthermore, Gibbs-Thomson-induced composition shift of the $\beta$ phase may result in a composition greater than 1. Both behaviors are non-physical if $c$ is the atomic fraction of solute, but can occur within the formulations in this paper because the free energy function is defined even for non-physical solute concentrations. To avoid these issues, the compositions of the $\alpha$ and $\beta$ phases are chosen as intermediate values within the atomic fraction interval, and the initial conditions presented in Sec. \ref{sub:ICs} are formulated such that the system will not exit the interval of $0\leq c \leq 1$ and $0\leq \eta \leq 1$.  

\subsection{Initial conditions, boundary conditions, and domain
geometries \label{sub:ICs}}

Several important factors were considered in determining the initial conditions and computational domains of the benchmark problems. First, the initial conditions for spinodal decomposition and precipitation simulations are typically created with a pseudorandom number generator.  However, the initial conditions must be repeatable from implementation to implementation in a benchmark problem, precluding the use of pseudorandom number generation.  Therefore, we choose trigonometric functions to provide smoothly varying, relatively disordered fields that are implementation-independent. Furthermore, the average composition and the amplitude and width of the fluctuations must be chosen such that phase separation will occur, as opposed to  the formation of a uniformly under- or supersaturated $\alpha$ phase.  Finally, the computational domain sizes and shapes are chosen
to stress the software implementation, because a wide variety of numerical
methods are currently in use. The domain sizes and interface resolution requirements must be large enough that runtime should be improved by parallel computing and mesh and
time adaptivity, yet not so large as to require significant resources on a high-performance computing cluster. We also anticipate that the use of non-rectilinear domains will become commonplace as new applications of the phase field method are investigated, such as nano-fabricated structures and cracking. Several phase field investigations (e.g., Refs. \cite{funkhouser2014dynamics, welland2015miscibility}) have already been performed with spherical domains.

Several boundary conditions, initial conditions, and computational domain geometries are used to challenge different aspects of the numerical solver implementation. For both benchmark problems, we test four combinations that are increasingly difficult to solve: two with square computational domains with side lengths of 200 units, one with a computational domain in the shape of a ``T'', with a total height of 120 units, a total width of 100 units, and horizontal and vertical section widths of 20 units (Fig.\ \ref{fig:p1_domains_IC}), and one in which the computational domain is the surface of a sphere with a radius of $r=100$ units. While most codes readily handle rectilinear domains, a spherical domain may pose problems, such as having the solution restricted to a two-dimensional curved surface. The coordinate systems and origins are given in Fig.\ \ref{fig:p1_domains_IC}. Periodic boundary conditions are applied to one square domain, while no-flux boundaries are applied to the other square domain and the ``T''-shaped domain. Periodic boundary conditions are commonly used with rectangular or rectangular prism domains to simulate an infinite material, while no-flux boundary conditions may be used to simulate an isolated piece of material or a mirror plane. As the computational domain is compact for the spherical surface, no boundary conditions are specified for it.  Note that the same initial conditions are used for the square computational domains with no-flux and periodic boundary conditions (Sections \ref{sub:spinodal_ICs} and \ref{sub:Ostwald_ICs}), such that when periodic boundary conditions are applied, there is a discontinuity in the initial condition at the domain boundaries.

\subsubsection{Spinodal decomposition \label{sub:spinodal_ICs}}

\begin{figure}
\begin{centering}
\subfloat[]{\begin{centering}
\includegraphics[scale=0.15]{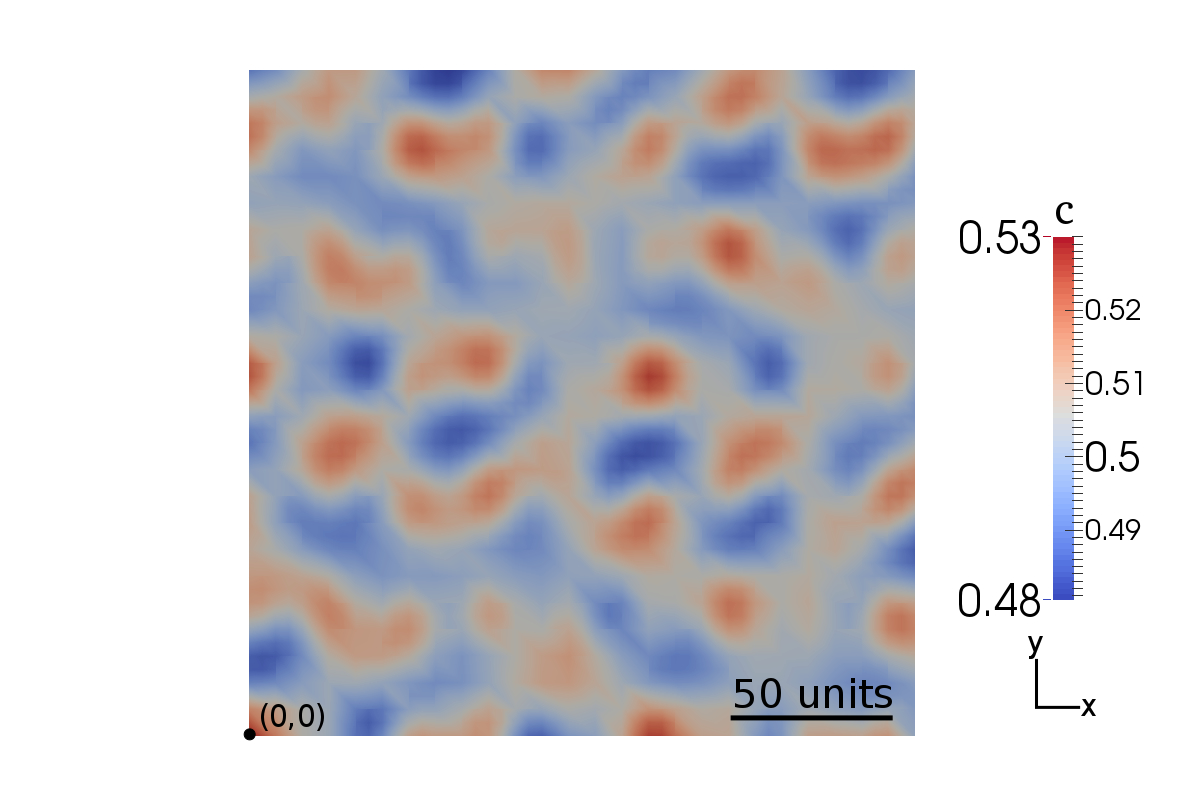}
\par\end{centering}

} \subfloat[]{\begin{centering}
\includegraphics[scale=0.15]{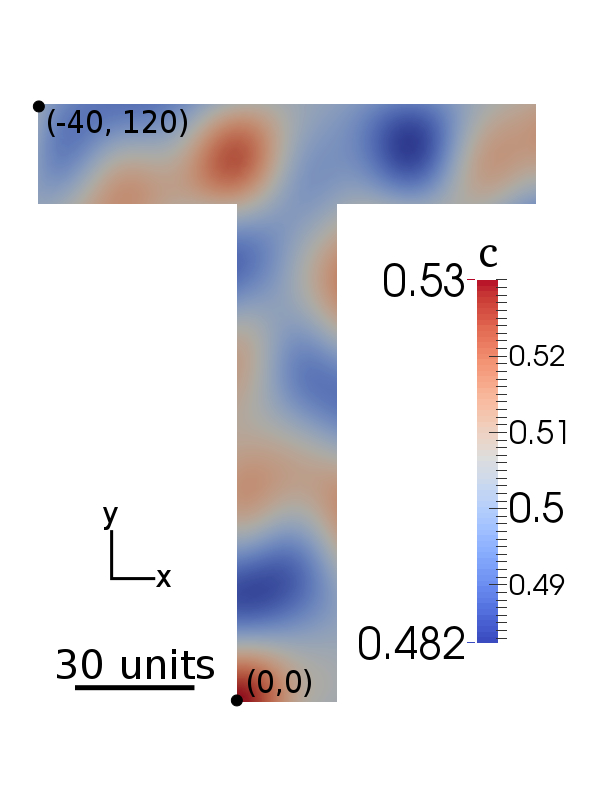}
\par\end{centering}

} \subfloat[]{\begin{centering}
\includegraphics[scale=0.15]{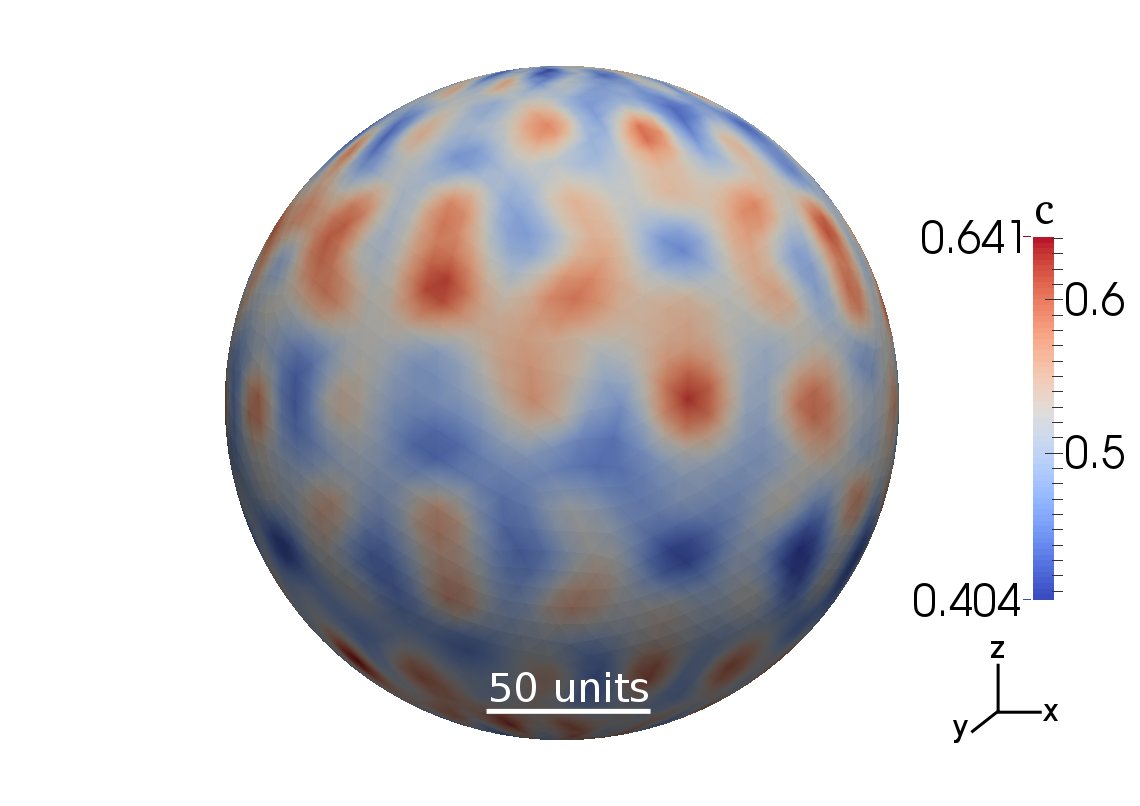}
\par\end{centering}

} 
\par\end{centering}

\caption{The computational domains and initial conditions for the spinodal decomposition benchmark problem. The origin for the coordinate system of the sphere is at its center. \label{fig:p1_domains_IC}}
\end{figure}

The initial conditions for the first benchmark problem are chosen such that the average value of $c$ over the computational domain is approximately $0.5$. The initial value of $c$ for the square and ``T'' computational domains is
specified by 
\begin{eqnarray}
c\left(x,y\right) & = & c_{0}+\epsilon\left[\cos\left(0.105x\right)\cos\left(0.11y\right)+\left[\cos\left(0.13x\right)\cos\left(0.087y\right)\right]^{2}\right.\nonumber \\
 &  & \left.+\cos\left(0.025x-0.15y\right)\cos\left(0.07x-0.02y\right)\right],\label{eq:p1_c_init}
\end{eqnarray}
where $c_{0}=0.5$ and $\epsilon=0.01$. In addition, the initial value of $c$ for the spherical computational domain is specified by
\begin{eqnarray}
c\left(\theta,\phi\right) & = & c_{0}+\epsilon_{sphere}\left[\cos\left(8\theta\right)\cos\left(15\phi\right)+\left(\cos\left(12\theta\right)\cos\left(10\phi\right)\right)^{2}\right.\nonumber \\
 &  & +\left.\cos\left(2.5\theta-1.5\phi\right)\cos\left(7\theta-2\phi\right)\right],\label{eq:p1_c_init_sp}
\end{eqnarray}
where $\epsilon_{sphere}=0.05$, and $\theta$ and $\phi$ are the polar and azimuthal angles, respectively, in a spherical coordinate system. These angles are translated into a Cartesian system as $\theta=\cos^{-1}\left(z/r\right)$ and $\phi=\tan^{-1}\left(y/x\right)$ dependent upon angle. The initial
conditions specified by Eqs.\ \ref{eq:p1_c_init} and \ref{eq:p1_c_init_sp} are shown in Fig.\ \ref{fig:p1_domains_IC}.

\subsubsection{Ostwald ripening \label{sub:Ostwald_ICs}}

\begin{figure}
\begin{centering}
\subfloat[]{\begin{centering}
\includegraphics[scale=0.15]{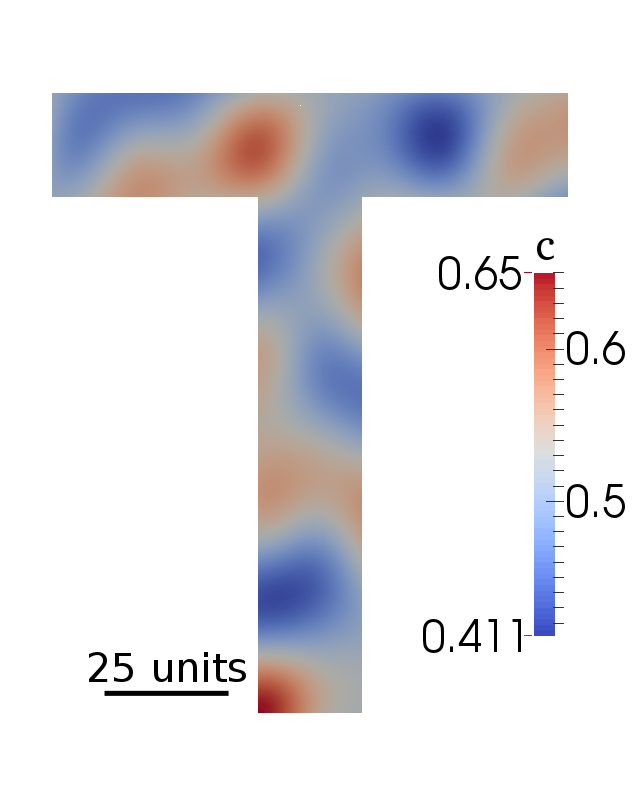}
\par\end{centering}

} \subfloat[]{\begin{centering}
\includegraphics[scale=0.15]{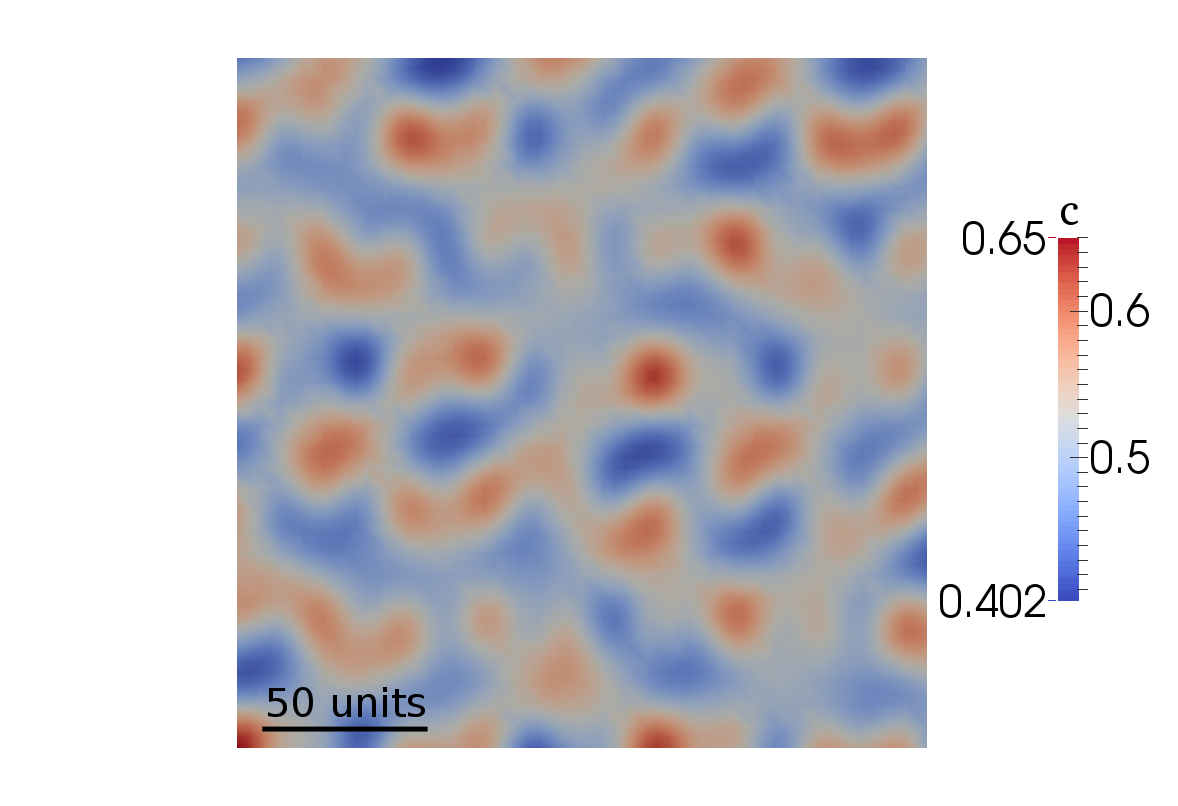}
\par\end{centering}

} \subfloat[]{\begin{centering}
\includegraphics[scale=0.15]{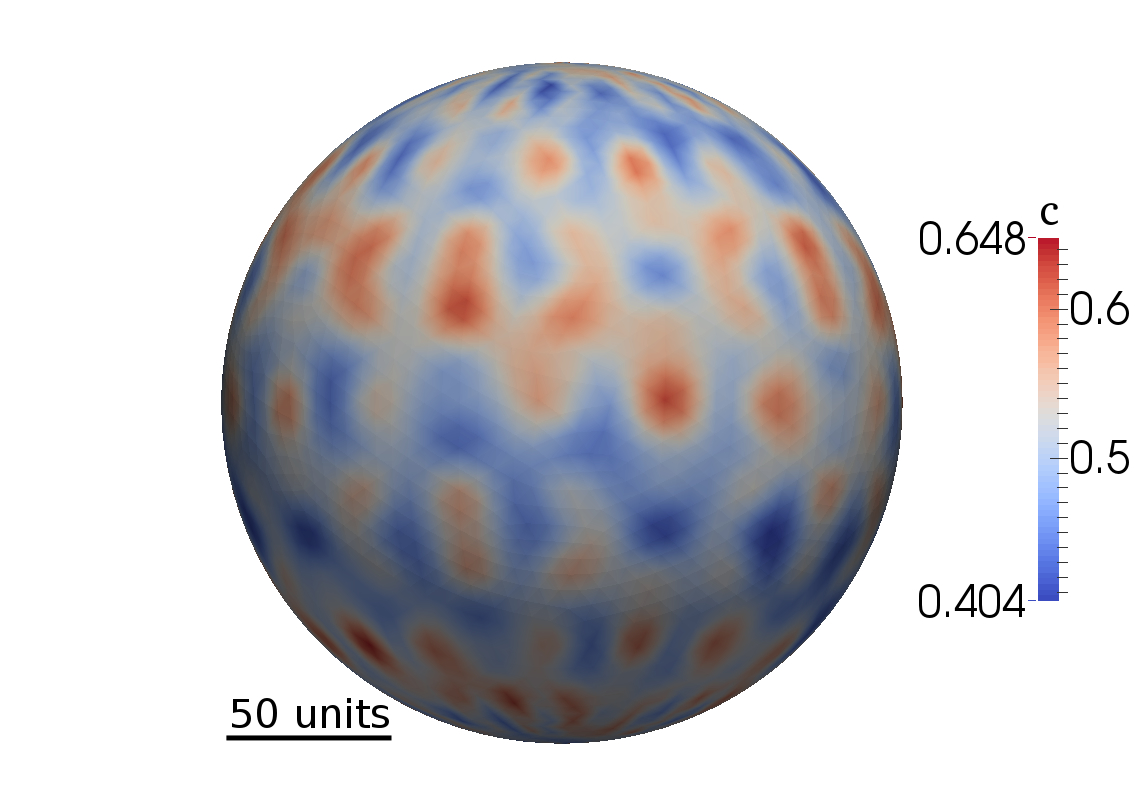}
\par\end{centering}

} 
\par\end{centering}

\begin{centering}
\subfloat[]{\begin{centering}
\includegraphics[scale=0.2]{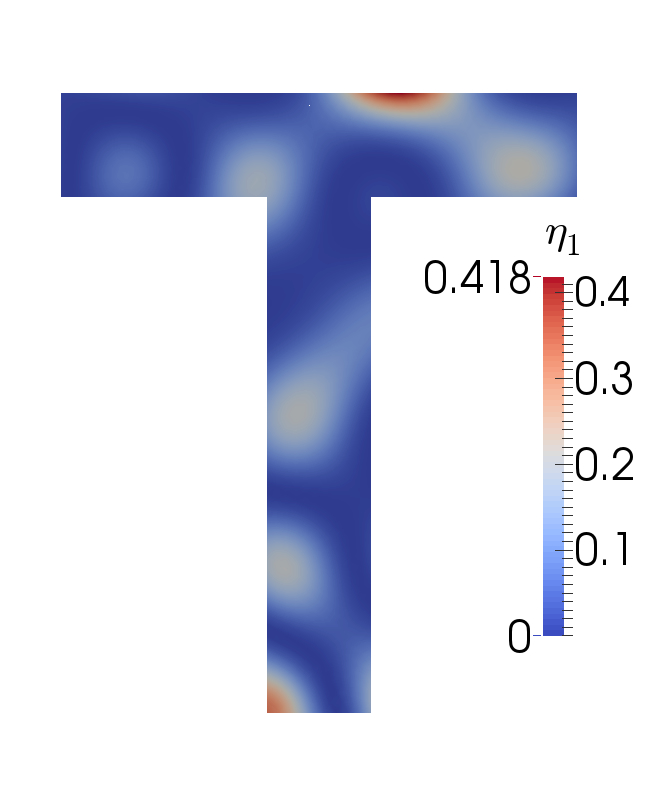}
\par\end{centering}

} \subfloat[]{\begin{centering}
\includegraphics[scale=0.2]{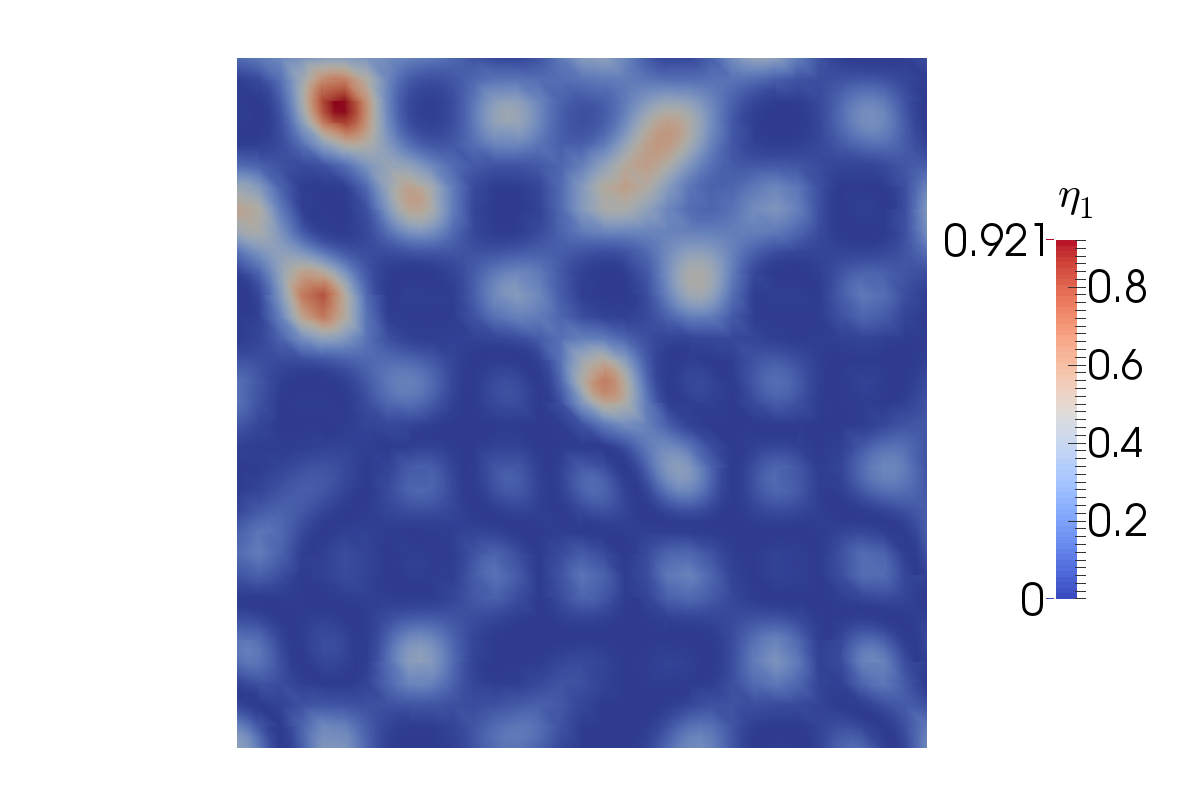}
\par\end{centering}

} \subfloat[]{\begin{centering}
\includegraphics[scale=0.2]{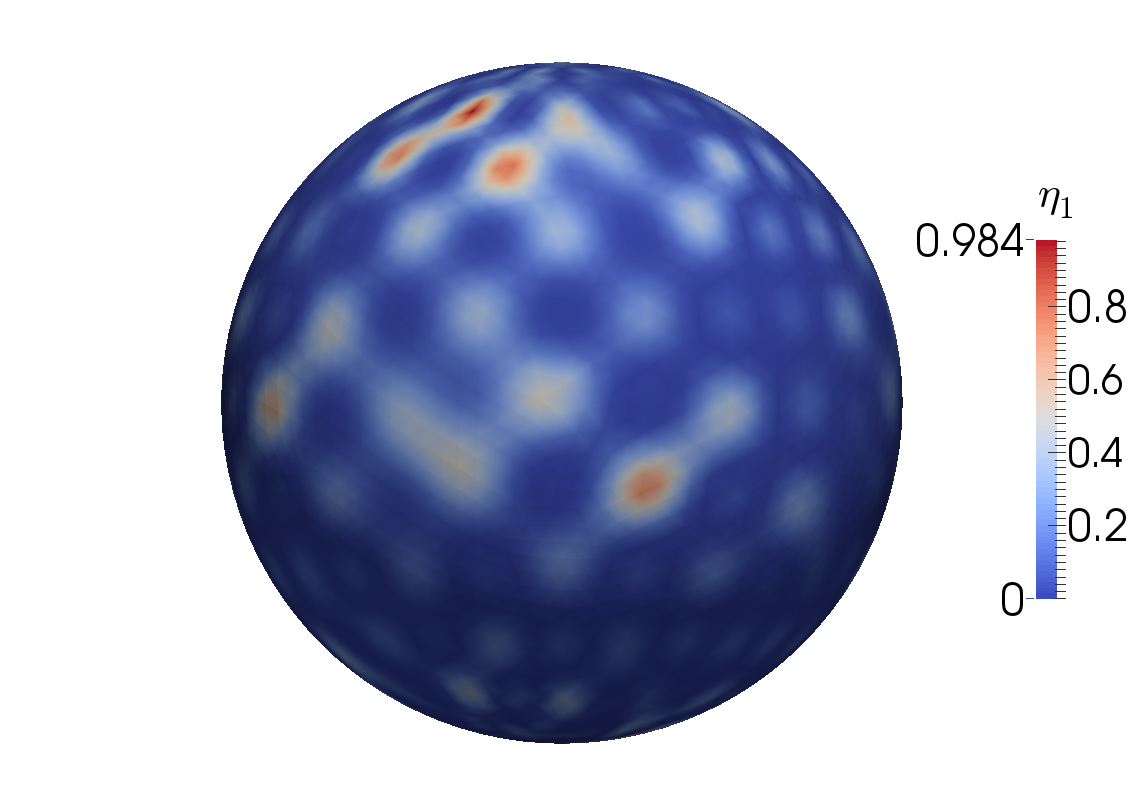}
\par\end{centering}

} 
\par\end{centering}

\begin{centering}
\subfloat[]{\begin{centering}
\includegraphics[scale=0.2]{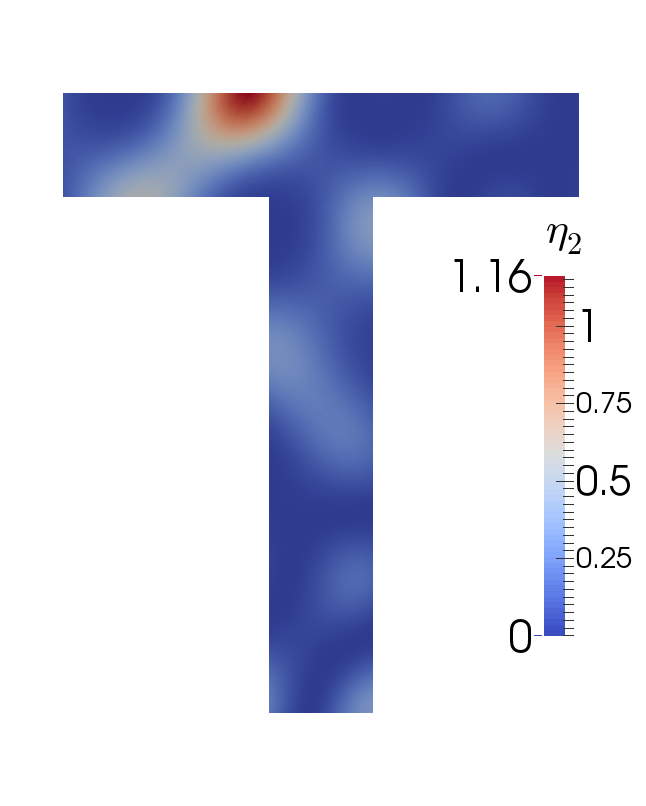}
\par\end{centering}

} \subfloat[]{\begin{centering}
\includegraphics[scale=0.2]{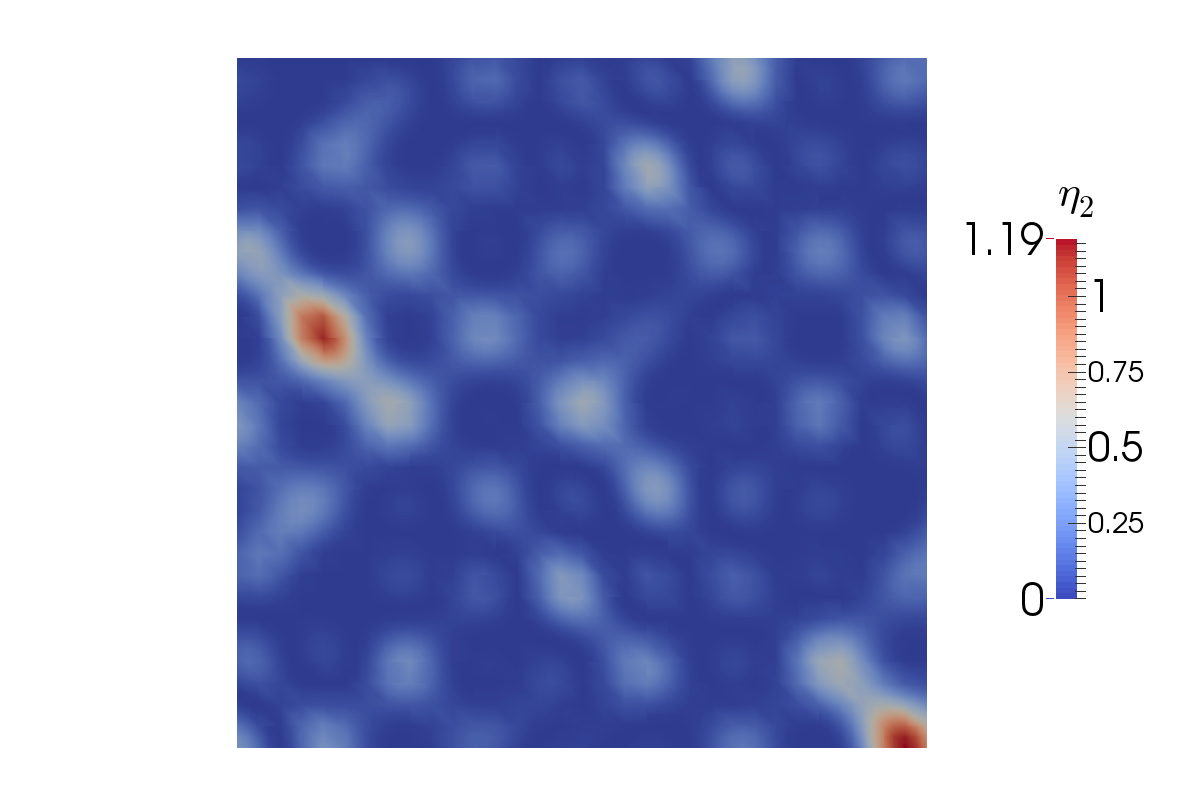}
\par\end{centering}

} \subfloat[]{\begin{centering}
\includegraphics[scale=0.2]{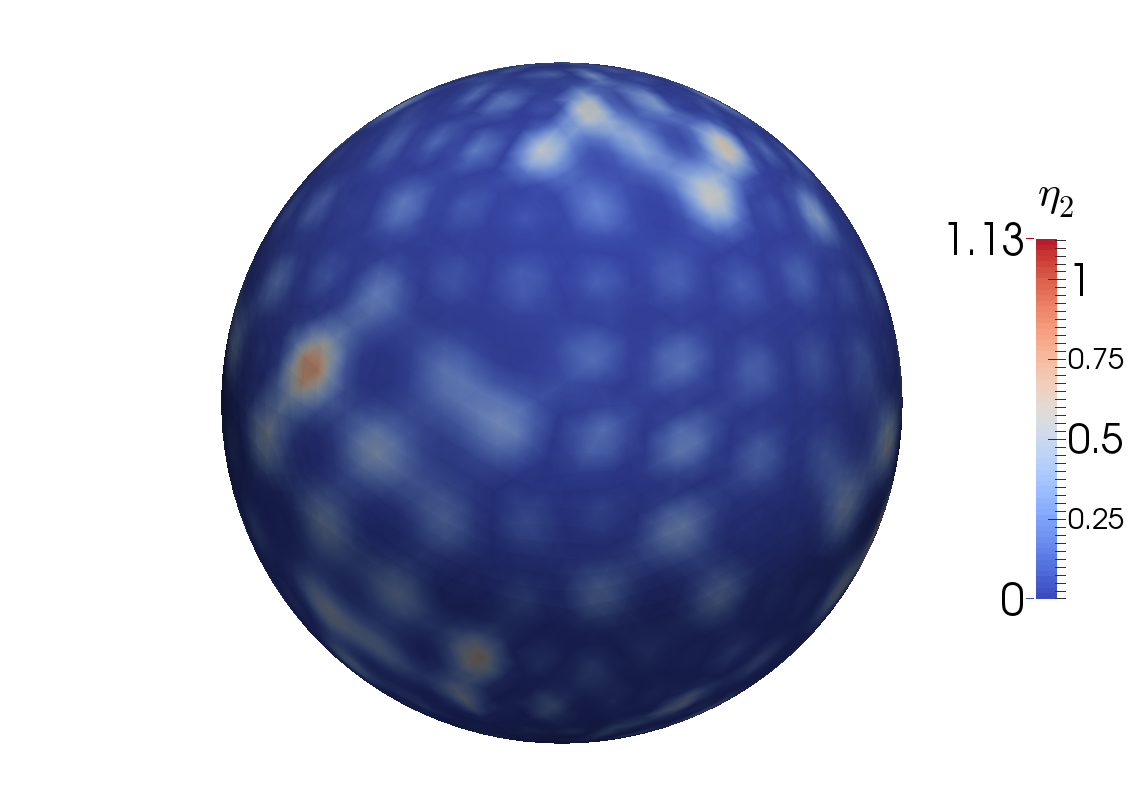}
\par\end{centering}

} 
\par\end{centering}

\caption{The computational domains and initial conditions for the Ostwald ripening benchmark problem. Top row: Initial conditions for the atomic fraction.
Middle row: Initial conditions for $\eta_{1}$. Bottom row: Initial conditions for $\eta_{2}$. Not shown: initial conditions for $\eta_{3}$ and $\eta_{4}$. \label{fig:p2_domains_ICs}}
\end{figure}

The initial conditions for the Ostwald ripening problem are qualitatively similar to those given for the spinodal decomposition problem, but the magnitude of the fluctuations around $c=0.5$ are greater, and fluctuations between [0,1] are applied to the structural order parameter fields. The initial condition for $c$ is again given by Eq.\ \ref{eq:p1_c_init} for the square and ``T'' domains, with $c_{0}=0.5$ and $\epsilon=0.05$, while it is given by Eq.\ \ref{eq:p1_c_init_sp} for the spherical domain, with $c_{0}=0.5$ and $\epsilon_{sphere}=0.05$. The initial condition for $\eta_{i}$ in the square and ``T'' domains is given as

\begin{eqnarray}
\eta_{i}\left(x,y\right) & = & \epsilon_{\eta}\left\{ \cos\left(\left(0.01i\right)x-4\right)\cos\left(\left(0.007+0.01i\right)y\right)\right.\nonumber \\
 &  & +\cos\left(\left(0.11+0.01i\right)x\right)\cos\left(\left(0.11+0.01i\right)y\right)\nonumber \\
 &  & +\psi\left[\cos\left(\left(0.046+0.001i\right)x+\left(0.0405+0.001i\right)y\right)\right.\nonumber \\
 &  & \left.\left.\cos\left(\left(0.031+0.001i\right)x-\left(0.004+0.001i\right)y\right)\right]^{2}\right\} ^{2},\label{eq:p2_n_init}
\end{eqnarray}
where $\epsilon_{\eta}=0.1$ and $\psi=1.5$, while for the
spherical domain, it is given as 
\begin{eqnarray}
\eta_{i}\left(\theta,\phi\right) & = & \epsilon_{\eta}^{sphere}\left\{ \cos\left(i\theta-4\right)\cos\left(\left(0.7+i\right)\phi\right)\right.\nonumber \\
 &  & +\cos\left(\left(11+i\right)\theta\right)\cos\left(\left(11+i\right)\phi\right)\nonumber \\
 &  & \psi\left[\cos\left(\left(4.6+0.1i\right)\theta+\left(4.05+0.1i\right)\phi\right)\right.\nonumber \\
 &  & \left.\left.\cos\left(\left(3.1+0.1i\right)\theta-\left(0.4+0.1i\right)\phi\right)\right]^{2}\right\} ^{2}\label{eq:p2_nsphere}
\end{eqnarray}
with $\epsilon_{\eta}^{sphere}=0.1$ and $i=1,\ldots,4$ enumerates the order parameters corresponding to the different phase variants.  The initial conditions for the Ostwald ripening simulations are shown
in Fig.~\ref{fig:p2_domains_ICs} for $c$, $\eta_{1}$, and $\eta_{2}$.

\section{Numerical methods}

To provide example solutions to the benchmark problems, the MOOSE computational framework is used. MOOSE \cite{gaston2014continuous,gaston2015physics} is an open-source finite element framework and is the basis for several other phase field applications, including Marmot \cite{tonks2012object} and Hyrax \cite{jokisaari2015general,jokisaari2016nucleation}. To avoid computationally expensive fourth-order derivative operators, the Cahn-Hilliard equation is split into the two second-order equations \cite{elliott1989second,tonks2012object}, given by 
\begin{equation}
\frac{\partial c}{\partial t}=\nabla\cdot\left(M\nabla\mu\right)\label{eq:CH_mu}
\end{equation}
and  
\begin{equation}
\mu=\frac{\partial f_{chem}}{\partial c}-\kappa\nabla^{2}c,\label{eq:mu}
\end{equation}
where $f_{chem}$ and $\kappa$ are given in Section \ref{sssec:spinodal_free_energy} and Section \ref{sssec:ostwald_free_energy} for the spinodal decomposition and Ostwald ripening problems, respectively.  
The square computational domains are meshed with square, four-node quadrilateral elements by the mesh generator within MOOSE, while the ``T''-shaped and spherical domains are meshed with triangular three-node elements using CUBIT \cite{CUBIT}.  Linear Lagrange shape functions are employed for $c$, $\eta_{i}$, and $\mu$. For computational efficiency, the system of nonlinear equations are solved with the full Newton method for the first problem, and the preconditioned Jacobian Free Newton-Krylov (PJFNK) method for the second problem.  The second backward differentiation formula (BDF2) \cite{iserles2009first} time integration scheme is applied in all cases. The simulations are solved with a nonlinear relative tolerance of $1\times10^{-8}$ and a nonlinear absolute tolerance of $1\times10^{-11}$.

To improve computational efficiency, adaptive meshing and adaptive time stepping are used. Each simulation is performed twice, once with the aggressive ``SolutionTimeAdaptive'' time stepper designed to finish the simulation as rapidly as possible  \cite{tonks2012object}, and once with the more conservative ``IterationAdaptive'' time stepper, which attempts to maintain a constant number of nonlinear iterations and a fixed ratio of nonlinear to linear iterations. We choose a target of five nonlinear iterations, plus or minus one, and a linear/nonlinear iteration ratio of 100. Both time adaptivity algorithms allow a maximum of 5\% increase per time step.  In addition, gradient jump indicators \cite{kirk2006libmesh} for $c$ and $\mu$ are used to determine mesh adaptivity, and the diffuse interface width spans at least five elements in all simulations.  

\section{Results and discussion}

In this section, we present lessons learned from the first Hackathon, the results of the two benchmark problems simulated with two different time adaptivity algorithms, and the needs that should be addressed with future benchmark problems. As discussed in the Introduction, many different software implementations exist for phase field models, including bespoke software developed in-house. While several phase field codes are designed to be applied to multiple types of problems, the possible multiphysics couplings are so varied that it may be impossible to develop a single phase field modeling framework to suit all phase field modeling needs. Benchmark problems will help the phase field community in assessing the accuracy and performance of individual software implementations.

Several lessons were learned from the first Hackathon hosted by CHiMaD, influencing the current benchmark problems as well as our design considerations for future problems. The Hackathon is a twenty-four hour event in which teams of two participants each use their phase field software of choice to simulate a specified set of phase field problems with whatever software and computational resources are available to them, including over the Internet. The goal of the Hackathon is to understand how different numerical implementations handle a set of phase field model problems of increasing difficulty with respect to accuracy and speed. We found that the original problem statements needed additional specifications for participants to successfully run the simulations without guesses or assumptions. Furthermore, the free energy functional, which was chosen from the literature, did not produce the phase compositions that were indicated. Finally, we needed standardized outputs for direct, quantitative comparison of the results.

\begin{figure}
\begin{centering}
\subfloat[]{\begin{centering}
\includegraphics[scale=0.9]{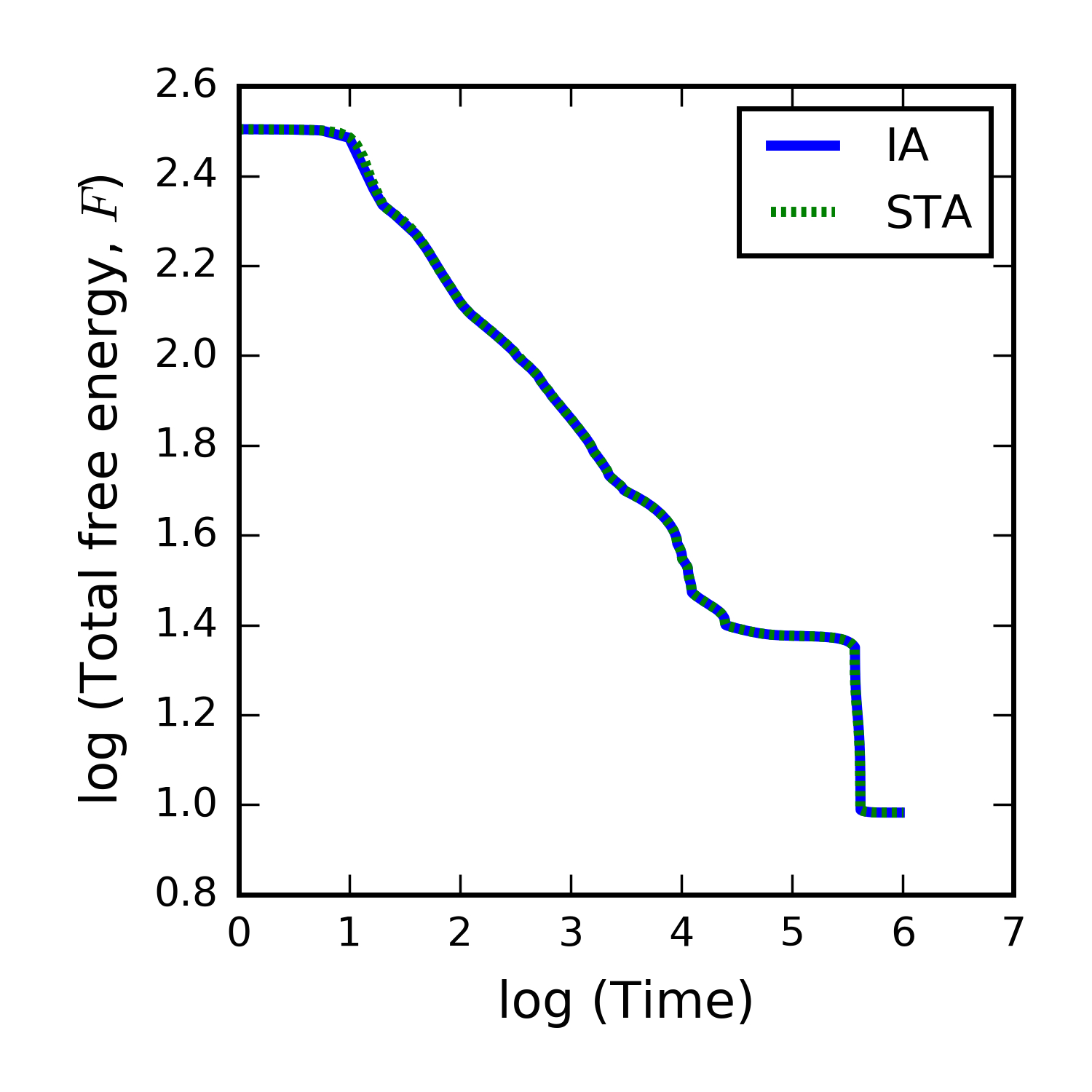}
\par\end{centering}

}\subfloat[]{\begin{centering}
\includegraphics[scale=0.9]{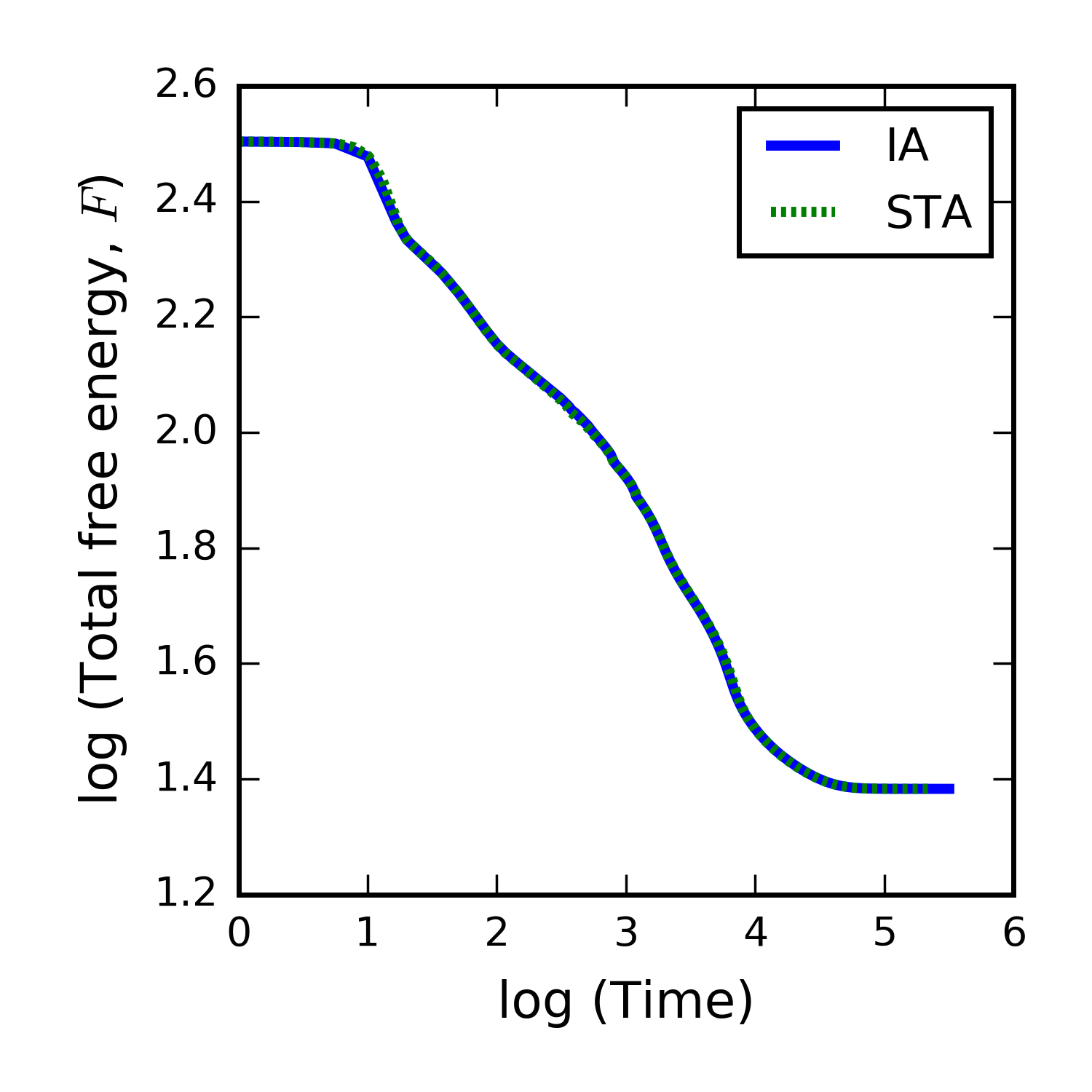}
\par\end{centering}

}
\par\end{centering}

\begin{centering}
\subfloat[]{\begin{centering}
\includegraphics[scale=0.9]{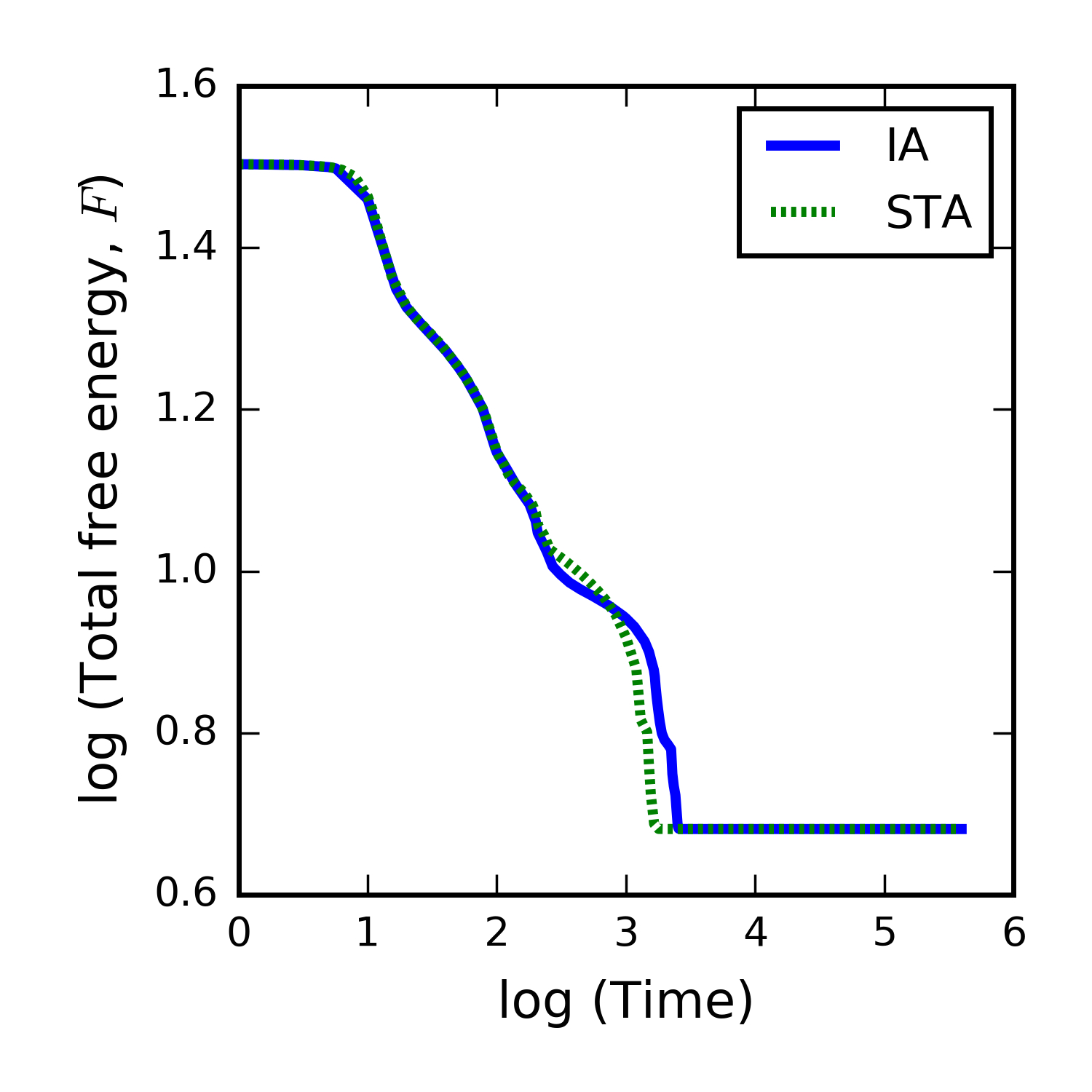}
\par\end{centering}

}\subfloat[]{\begin{centering}
\includegraphics[scale=0.9]{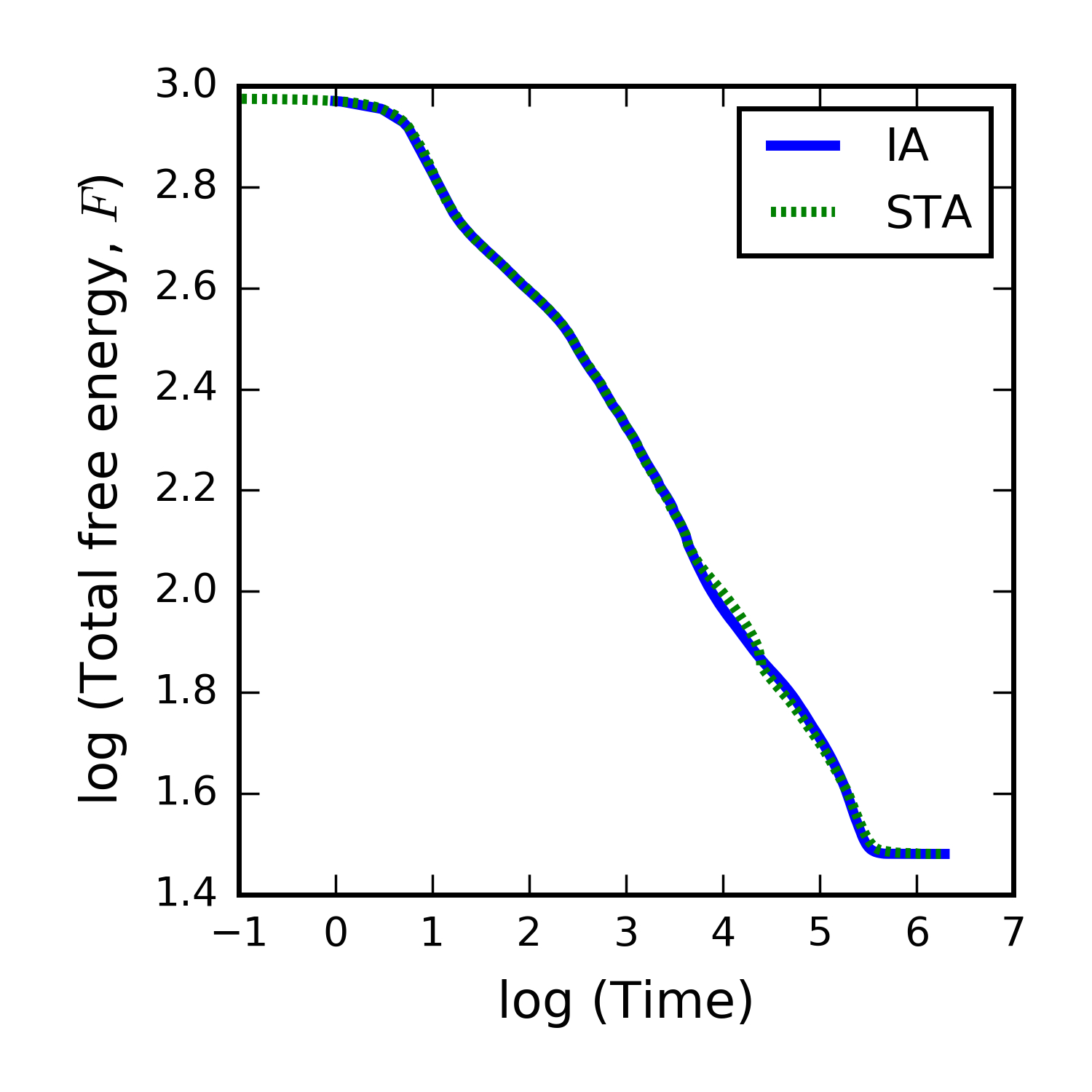}
\par\end{centering}

} 
\par\end{centering}

\caption{The total free energy evolution of the different variations of the spinodal decomposition benchmark problem simulated with two different time steppers, for (a) the square computational domain with no-flux boundary conditions, (b) the square computational domain with periodic boundary conditions, (c) the T-shaped computational domain, and (d) the spherical surface domain. ``IA'' indicates the conservative ``IterationAdaptive'' time stepper within MOOSE, and ``STA'' indicates the aggressive ``SolutionTimeAdaptive'' time stepper. \label{fig:p1_energy_evolution}}
\end{figure}

\begin{figure}
\begin{centering}
\subfloat[]{\begin{centering}
\includegraphics[scale=0.9]{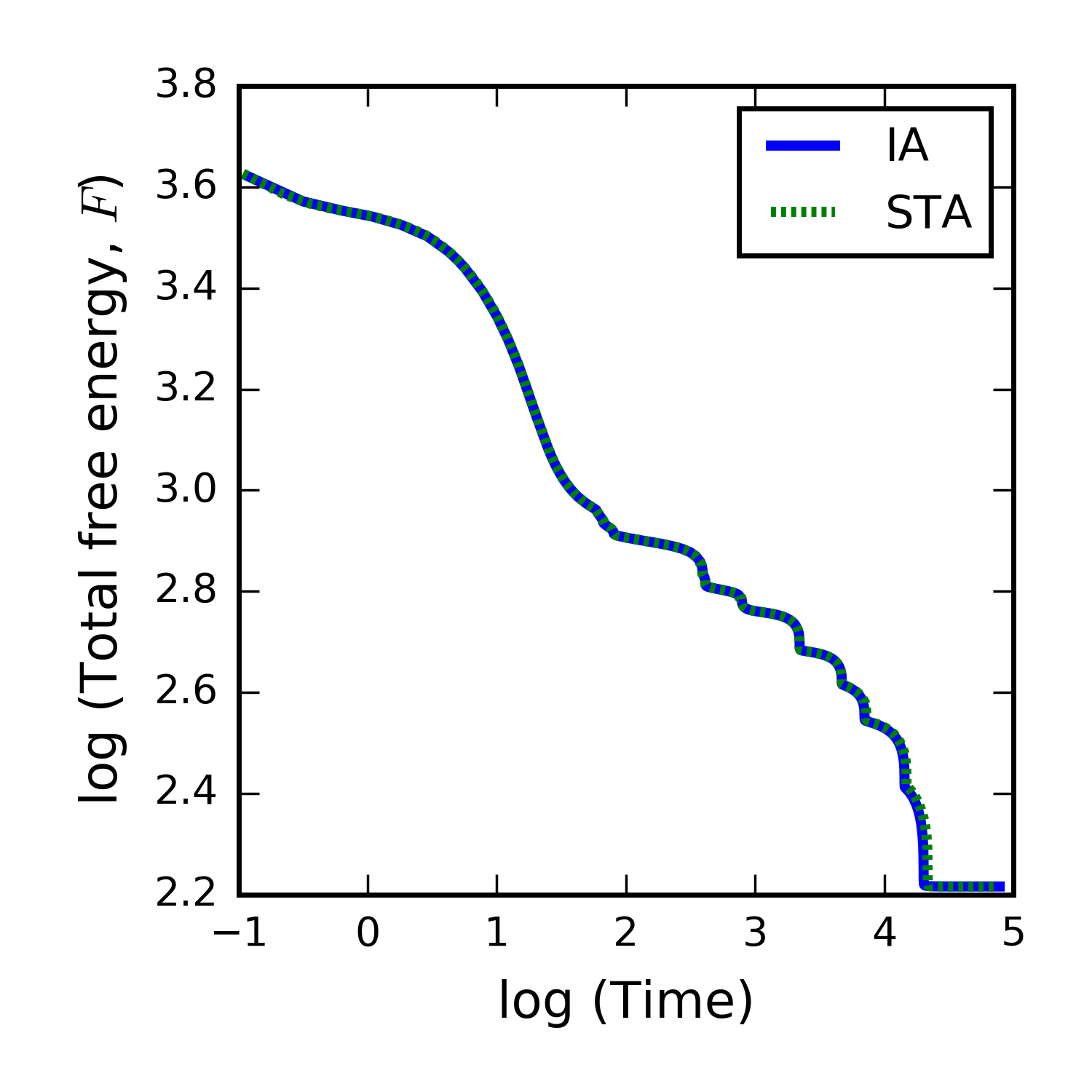}
\par\end{centering}

}\subfloat[]{\begin{centering}
\includegraphics[scale=0.9]{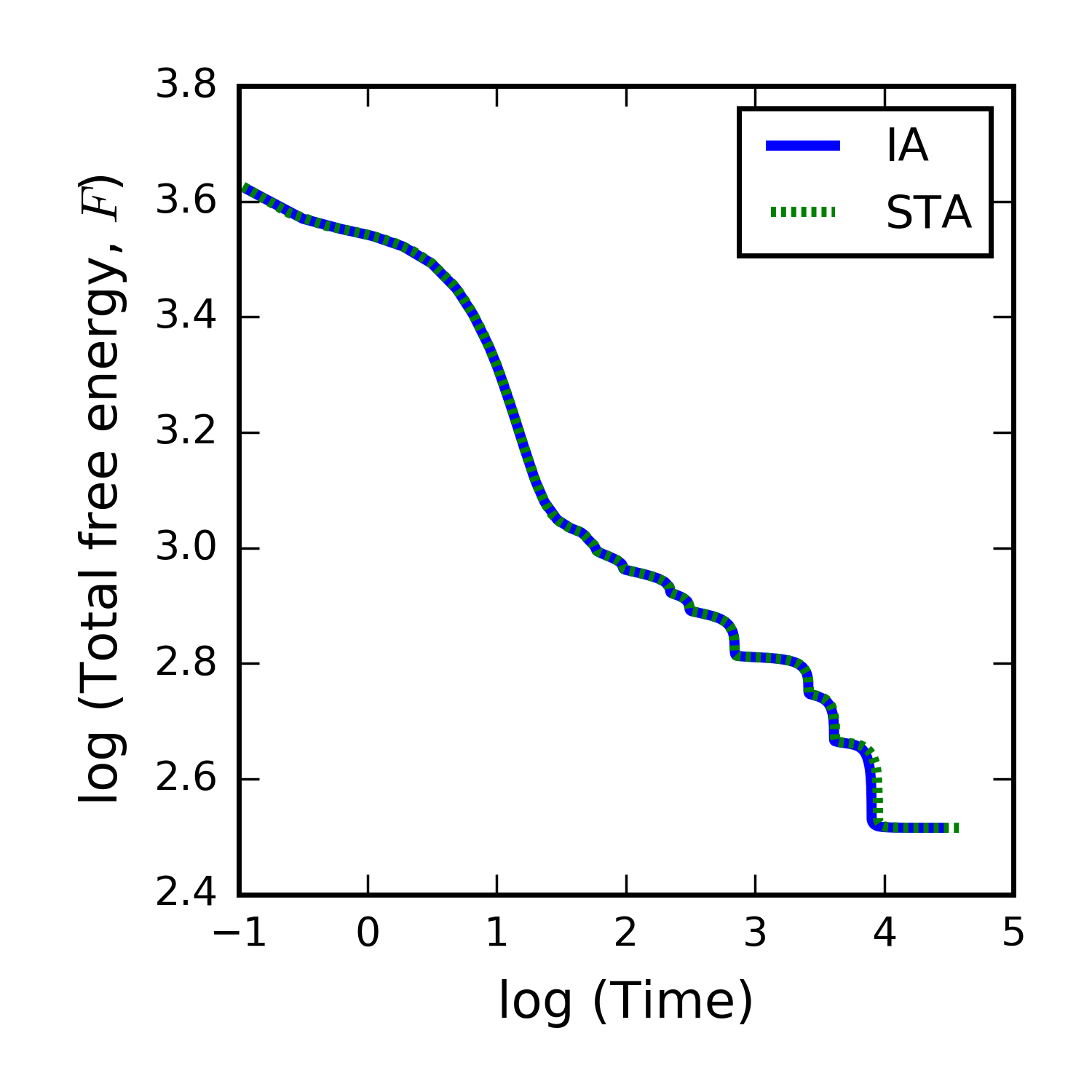}
\par\end{centering}

}
\par\end{centering}

\begin{centering}
\subfloat[]{\begin{centering}
\includegraphics[scale=0.9]{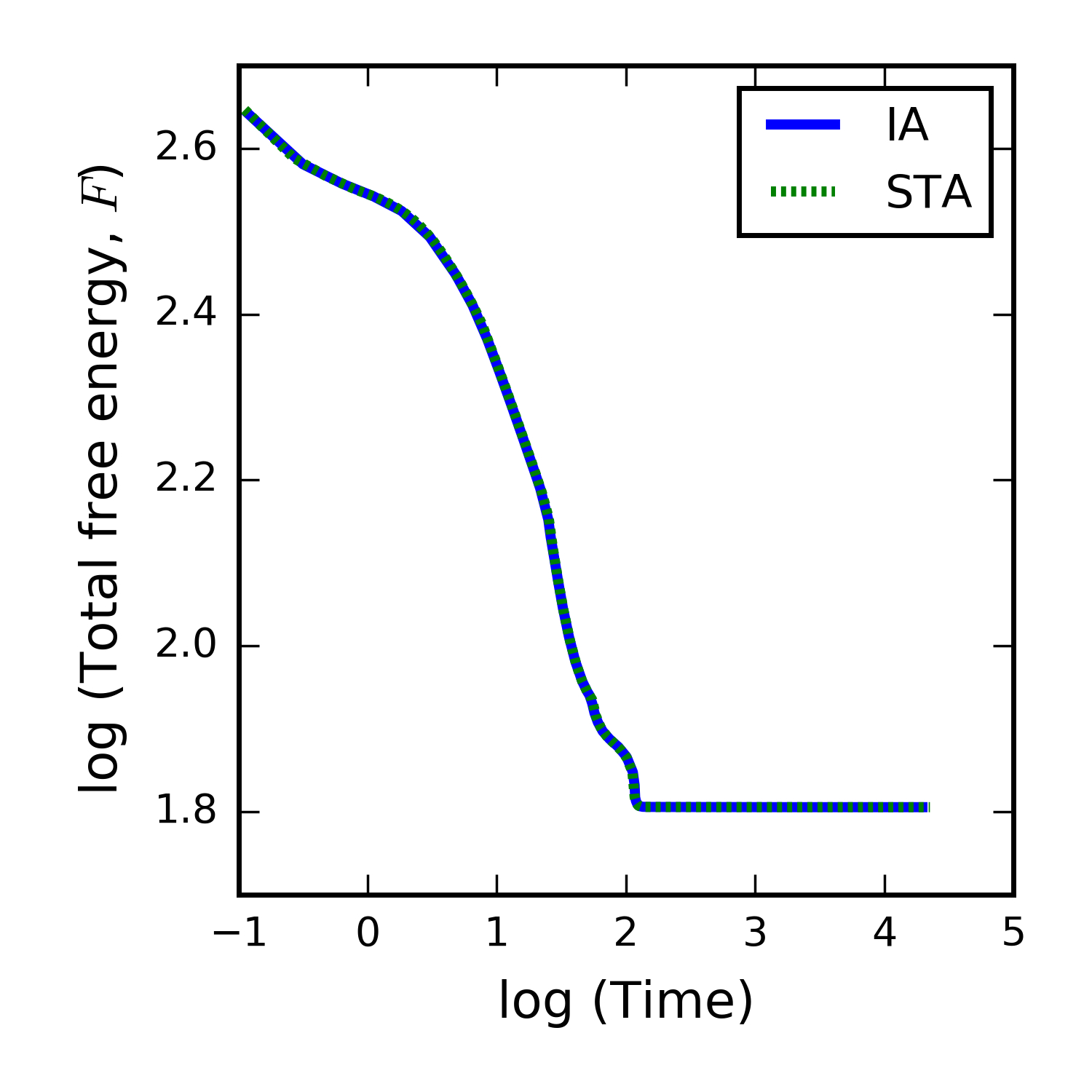}
\par\end{centering}

}\subfloat[]{\begin{centering}
\includegraphics[scale=0.9]{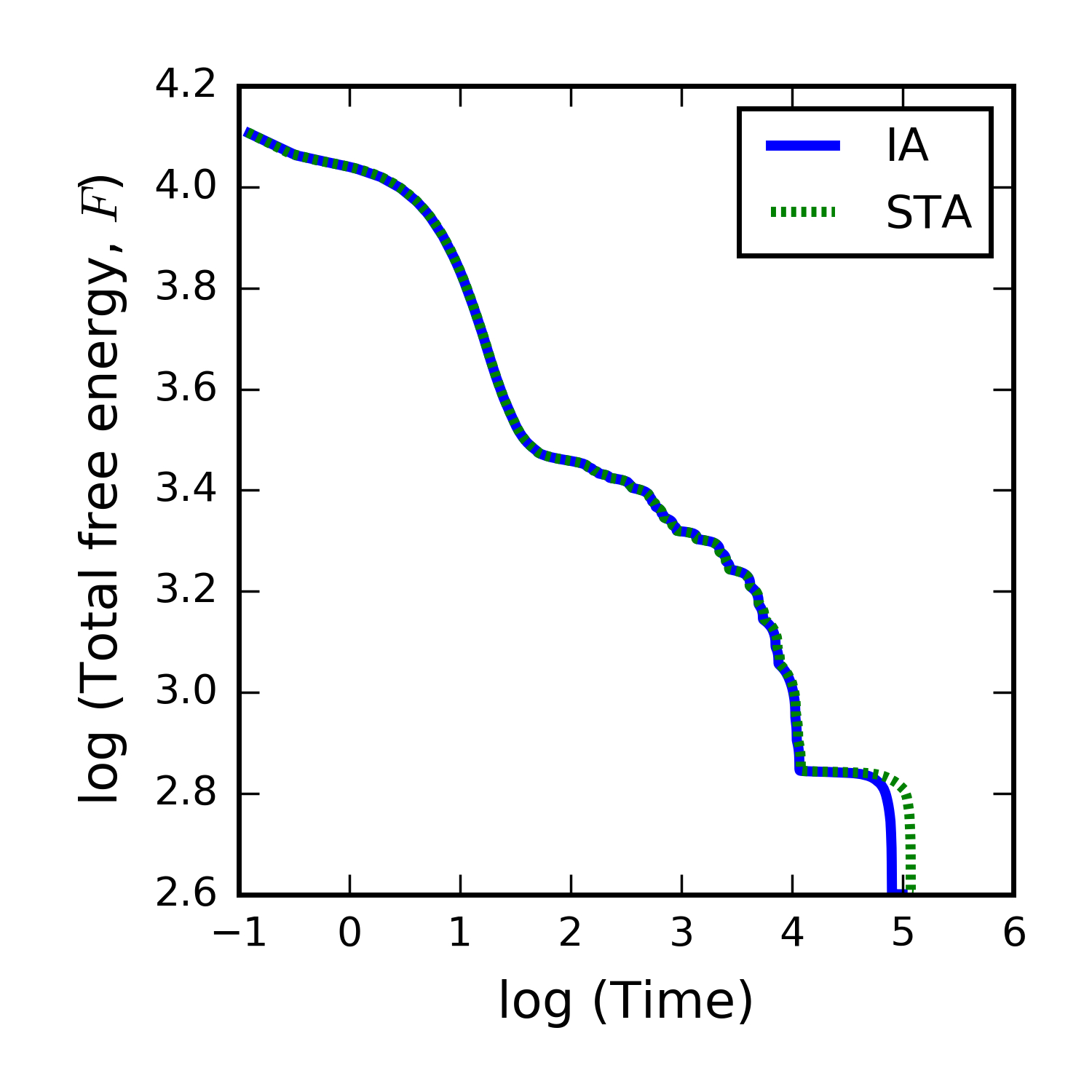}
\par\end{centering}

} 
\par\end{centering}

\caption{The total free energy evolution of the different variations of the Ostwald ripening benchmark problem simulated with two different time steppers, for (a) the square computational domain with no-flux boundary conditions, (b) the square computational domain with periodic boundary conditions, (c) the T-shaped computational domain, and (d) the spherical surface domain. ``IA'' indicates the conservative ``IterationAdaptive'' time stepper within MOOSE, and ``STA'' indicates the aggressive ``SolutionTimeAdaptive'' time stepper.  \label{fig:p2_energy_evolution}}
\end{figure}

\begin{figure}

\begin{centering}
\subfloat[]{\begin{centering}
\includegraphics[scale=0.135]{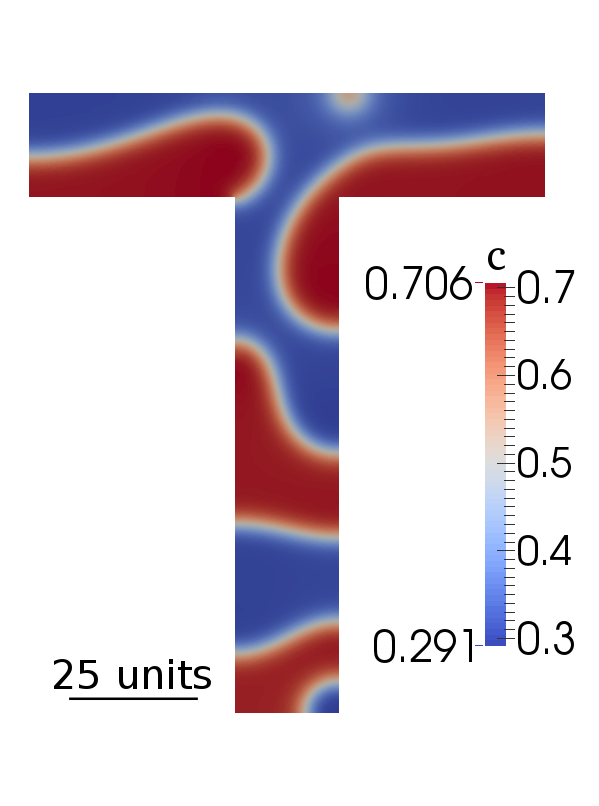}
\par\end{centering}

}\subfloat[]{\begin{centering}
\includegraphics[scale=0.18]{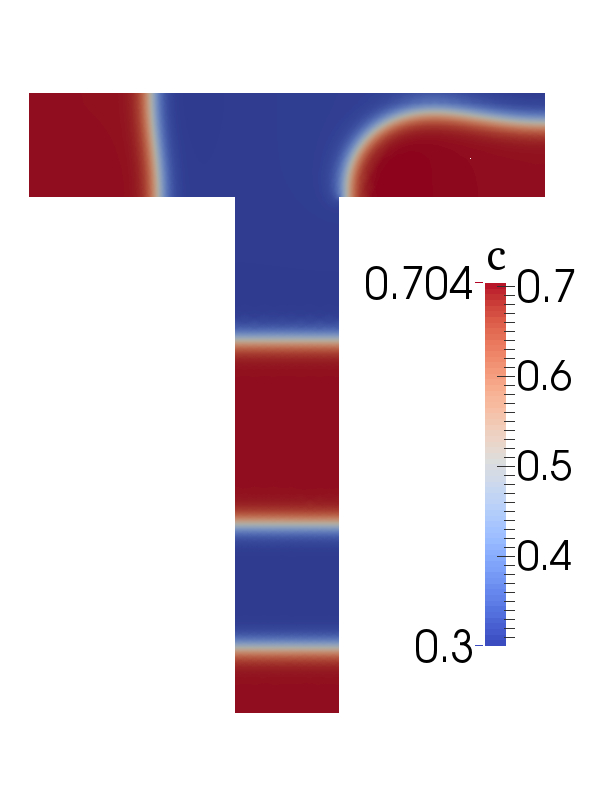}
\par\end{centering}

}\subfloat[]{\begin{centering}
\includegraphics[scale=0.18]{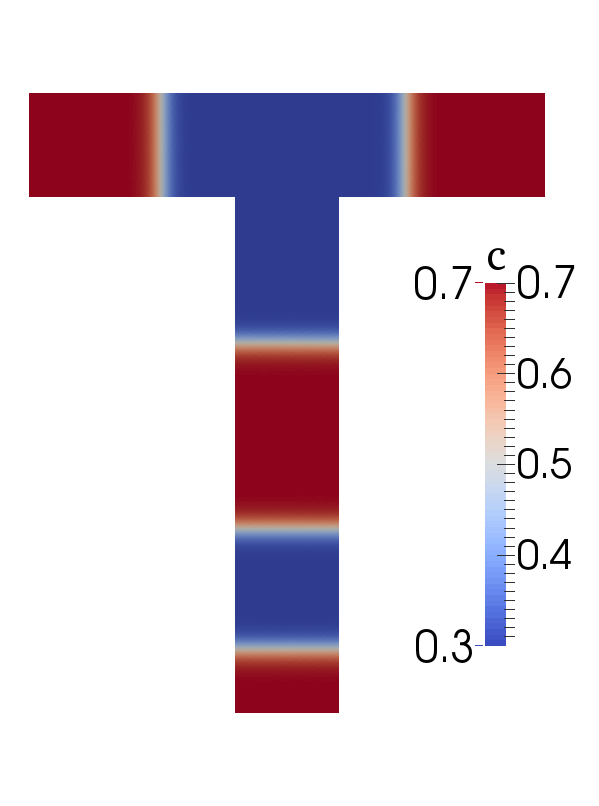}
\par\end{centering}
}
\par\end{centering}

\begin{centering}
\subfloat[]{\begin{centering}
\includegraphics[scale=0.18]{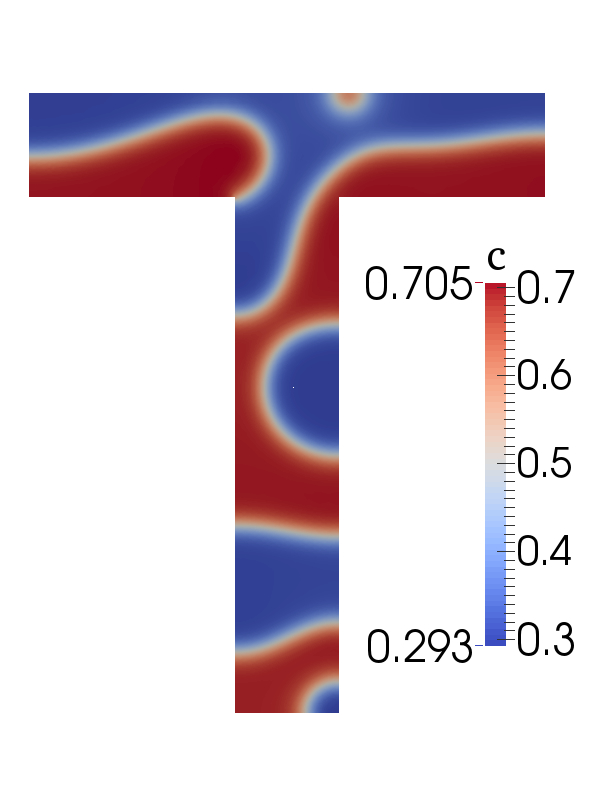}
\par\end{centering}

}\subfloat[]{\begin{centering}
\includegraphics[scale=0.18]{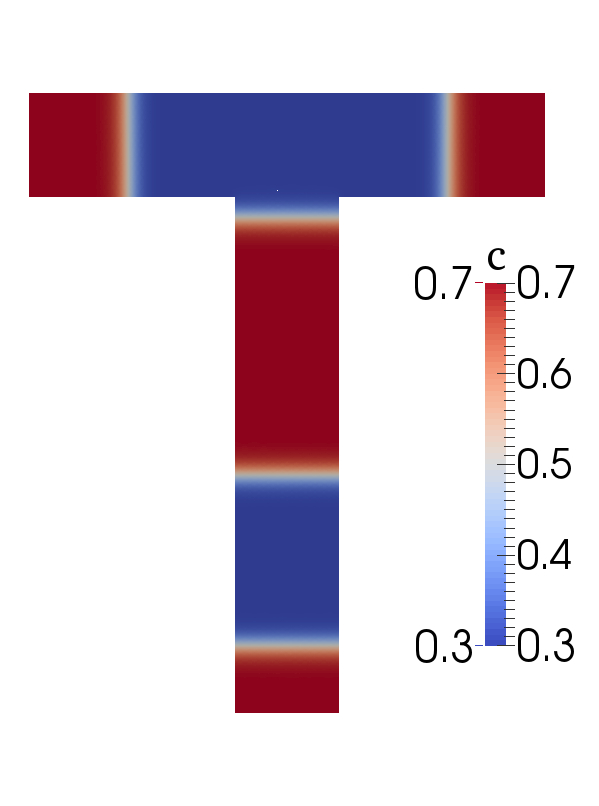}
\par\end{centering}

}\subfloat[]{\begin{centering}
\includegraphics[scale=0.18]{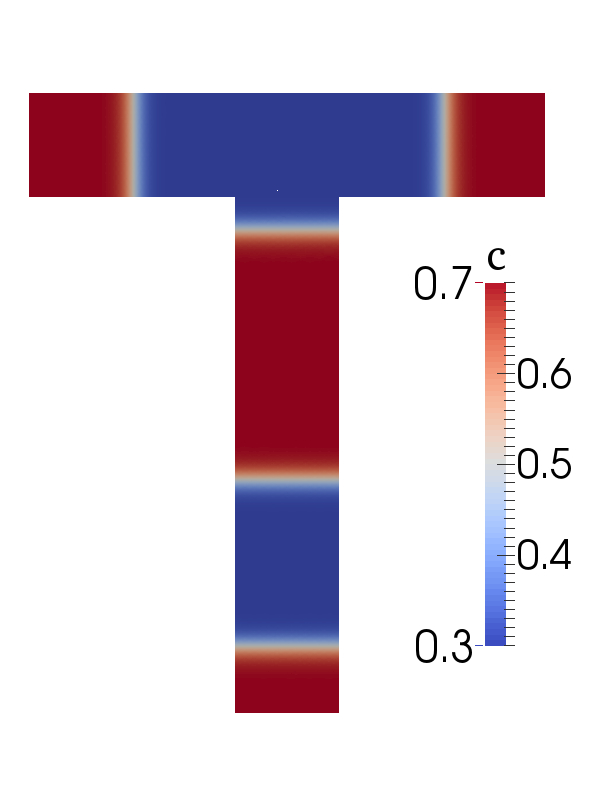}
\par\end{centering}
}
\par\end{centering}

\begin{centering}
\subfloat[]{\begin{centering}
\includegraphics[scale=0.115]{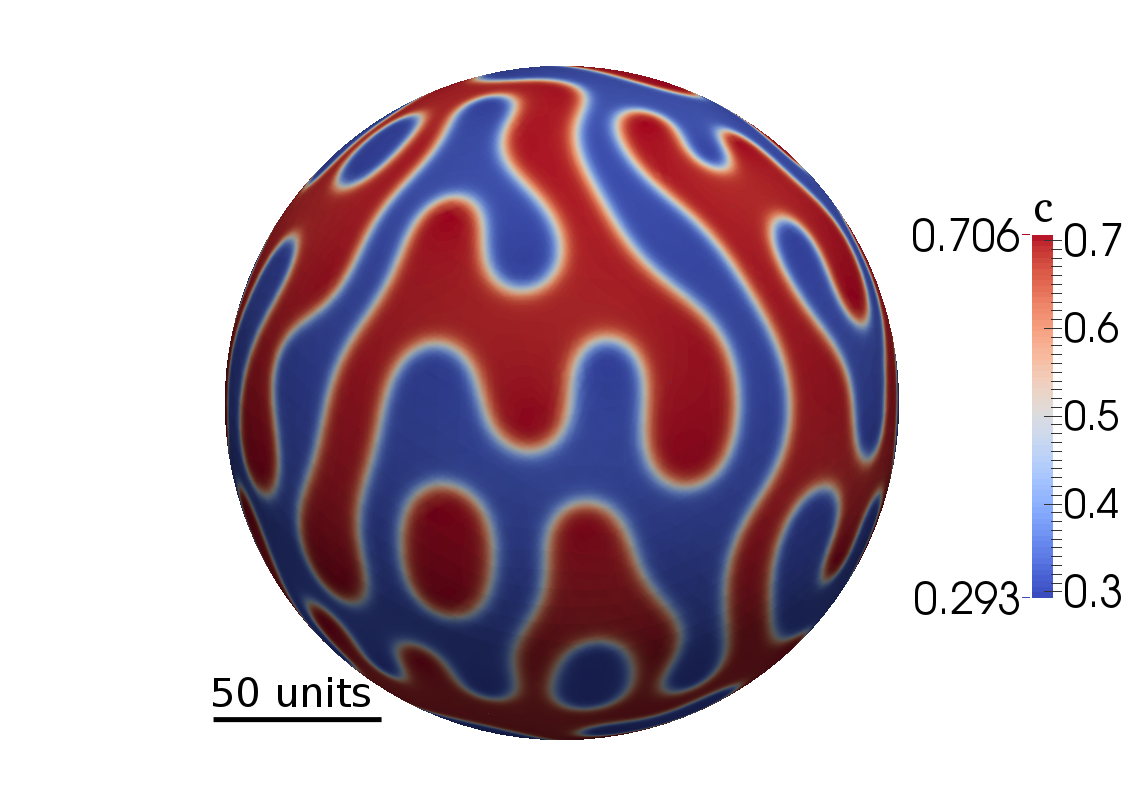}
\par\end{centering}

}\subfloat[]{\begin{centering}
\includegraphics[scale=0.15]{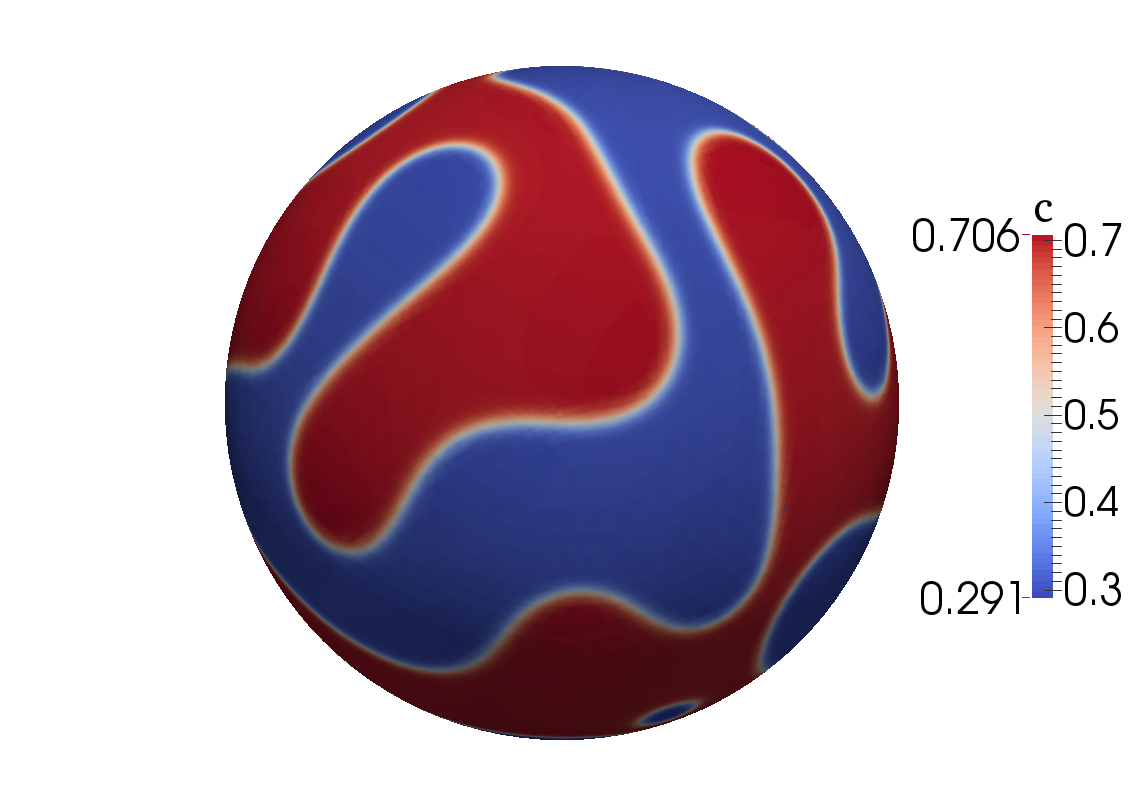}
\par\end{centering}

}\subfloat[]{\begin{centering}
\includegraphics[scale=0.15]{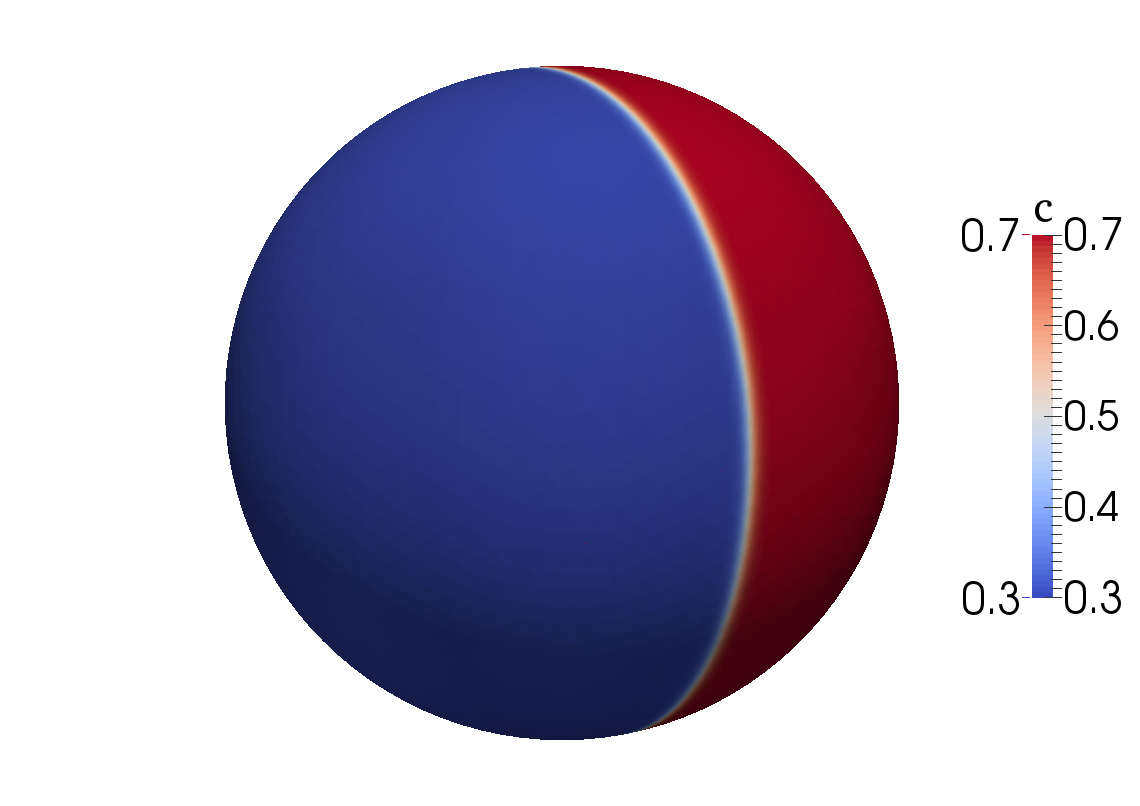}
\par\end{centering}
}
\par\end{centering}

\begin{centering}
\subfloat[]{\begin{centering}
\includegraphics[scale=0.15]{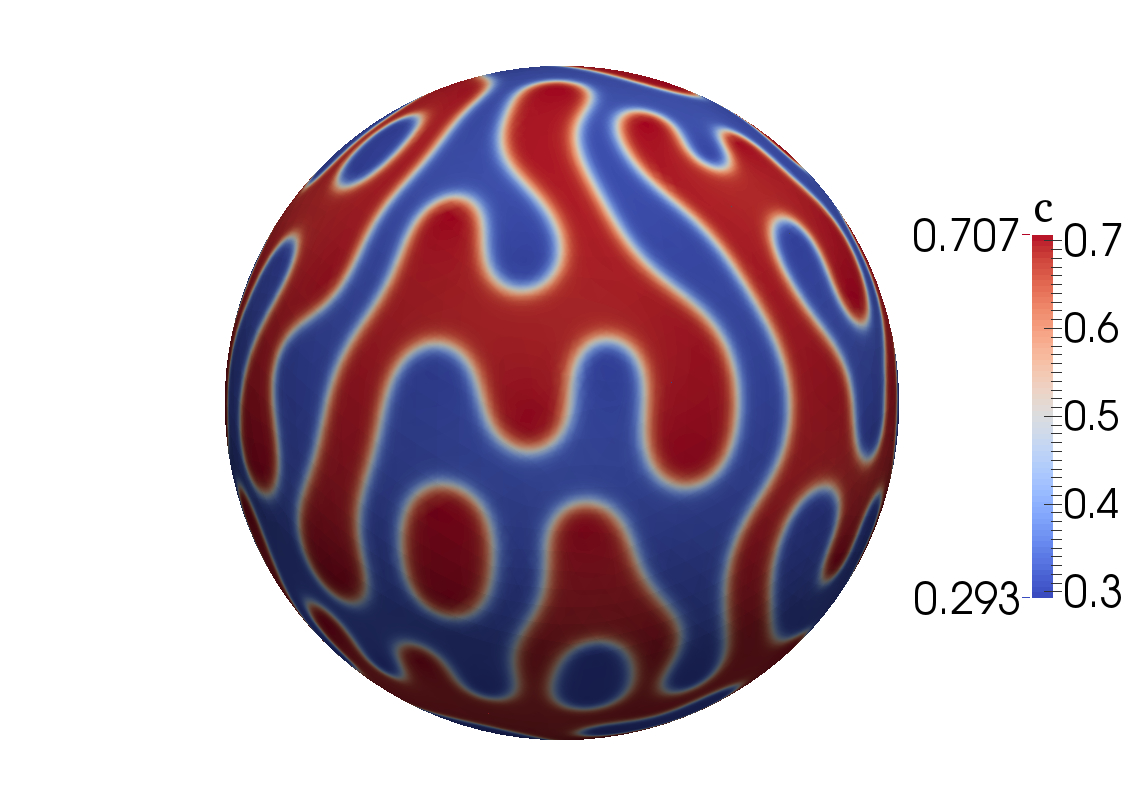}
\par\end{centering}

}\subfloat[]{\begin{centering}
\includegraphics[scale=0.15]{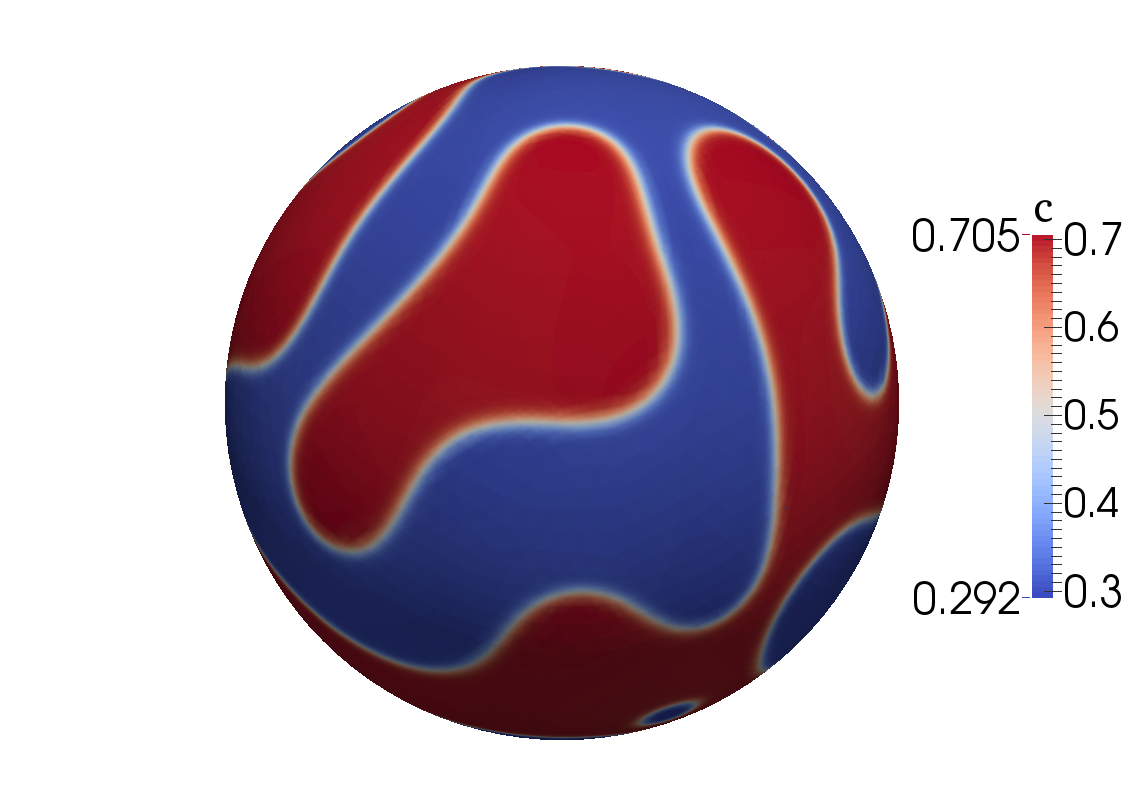}
\par\end{centering}

}\subfloat[]{\begin{centering}
\includegraphics[scale=0.15]{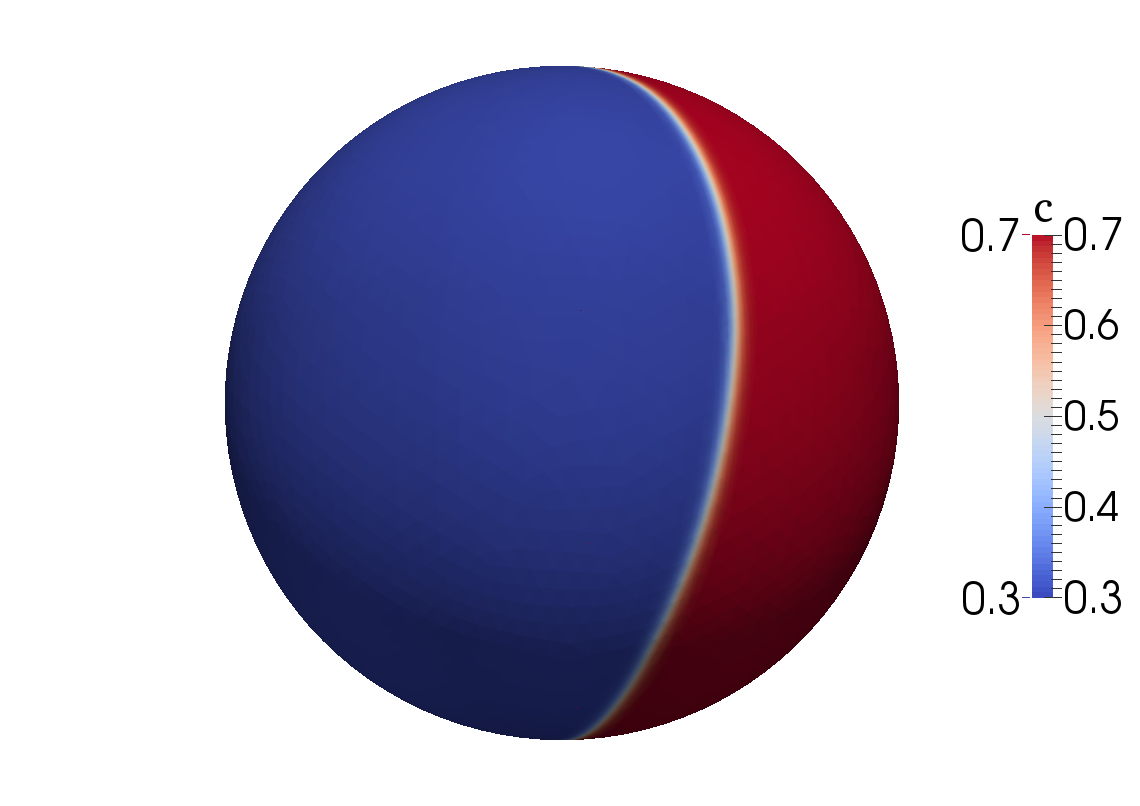}
\par\end{centering}
 
} 
\par\end{centering}

\caption{Snapshots of the microstructure evolution for spinodal decomposition simulated with different time steppers.  (a-c), (g-i): conservative time stepper; (d-f), (j-l): aggressive time stepper.  (a), (d), (g), (j): $t=200$; (b), (e), (h), (k): $t=2000$; (c), (f), (i), (l): end time. Note the clearly visible differences in microstructure between the two time steppers in (b) and (e), and (h) and (k).\label{fig:p1_microstructure}}
\end{figure}

For these benchmark problems, we choose the total free energy of the system and microstructural snapshots as the metrics to compare simulation results. Because we use time adaptivity, we choose several synchronization times ($t=$ 1, 5, 10, 20, 100, 200, 500, 1000, 2000, 3000, and 10000) so that simulation results obtained from the different time steppers may be directly compared at given times. Figures \ref{fig:p1_energy_evolution} and \ref{fig:p2_energy_evolution} show the total free energy of the different simulations of spinodal decomposition and Ostwald ripening, respectively. In all cases, the total free energy decreases rapidly, then asymptotically approaches the local energy minimum of the system (which varies given the initial and boundary conditions).  While the starting and final free energies are the same for each set of simulations (e.g., spinodal decomposition in the square domain with no-flux boundary conditions), the evolution of the energy is affected by the choice of the adaptive time stepper.  In the spinodal decomposition problem, the differences in energy as a result of the different time steppers are small for the square computational domains, but are more significant for the spherical and T-shaped computational domains. As shown in Fig.\ \ref{fig:p1_microstructure}, which presents microstructure snapshots for spinodal decomposition at $t=200$,  $t=2000$, and the end of the simulation, obvious microstructural differences are discernible.  Small variations early in the simulations strongly affect the microstructural evolution at later times, even in some cases affecting the final lowest-energy structure, as seen for the T-shaped computational domain. 

\begin{figure}

\begin{centering}
\subfloat[]{\begin{centering}
\includegraphics[scale=0.19]{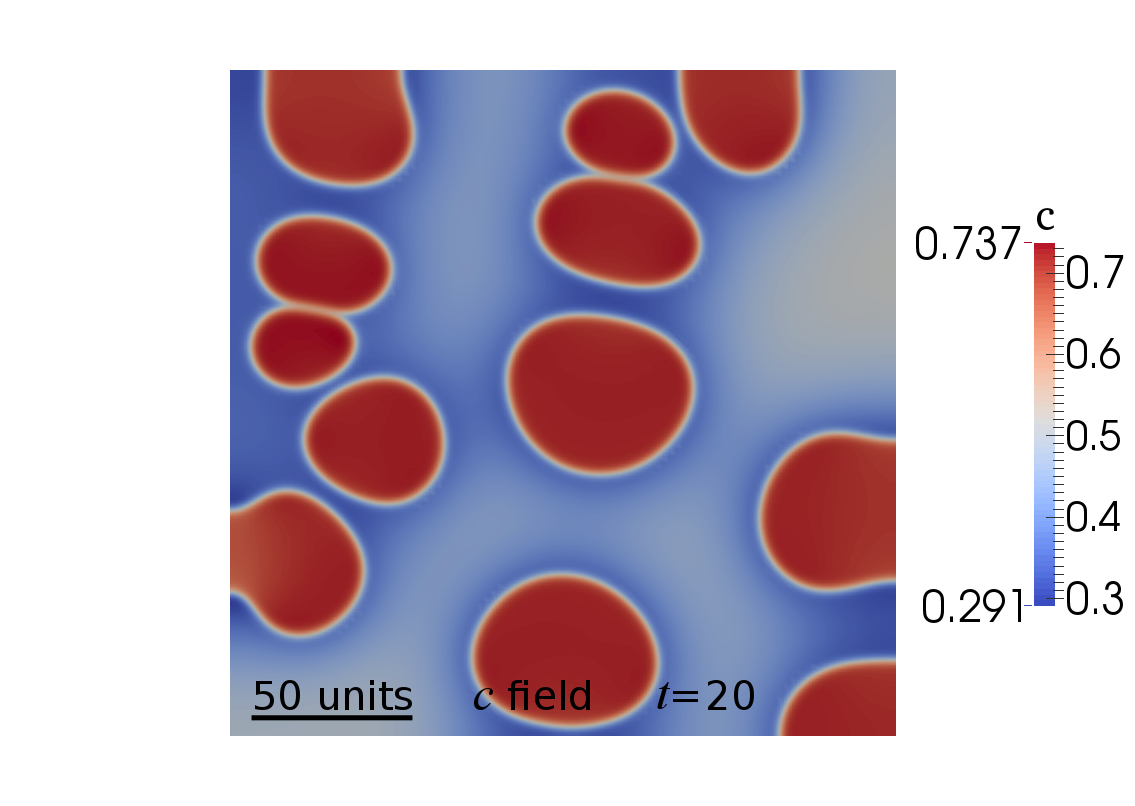}
\par\end{centering}

}\subfloat[]{\begin{centering}
\includegraphics[scale=0.19]{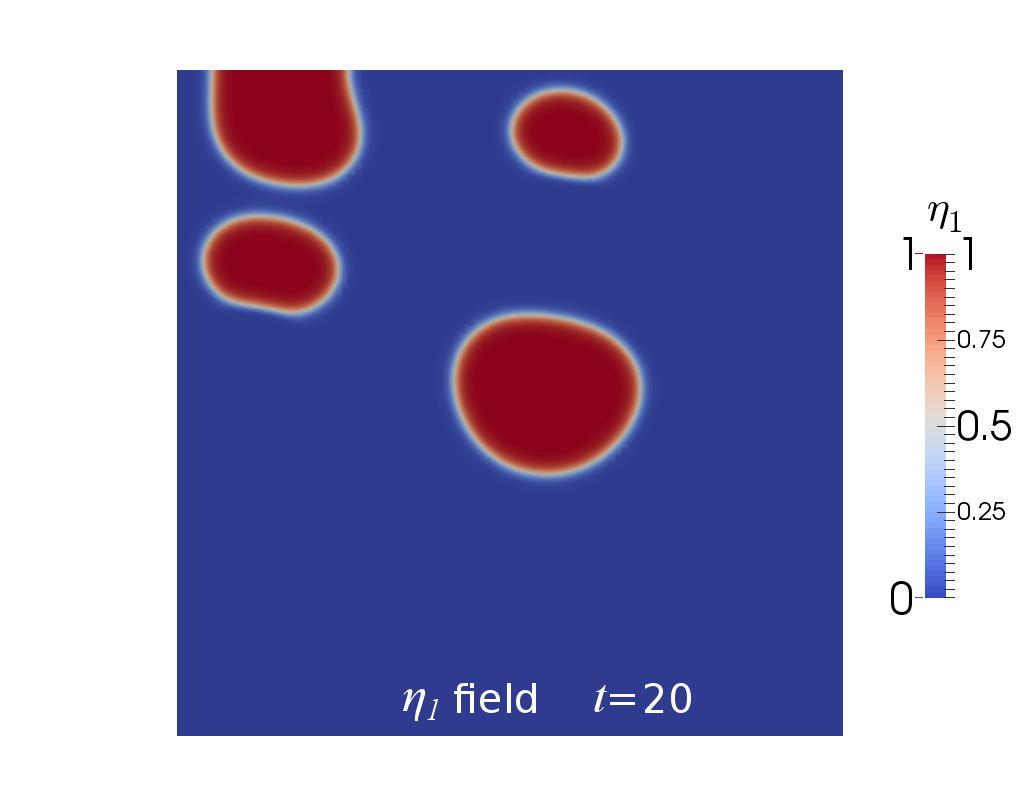}
\par\end{centering}

}\par\end{centering}

\begin{centering}
\subfloat[]{\begin{centering}
\includegraphics[scale=0.19]{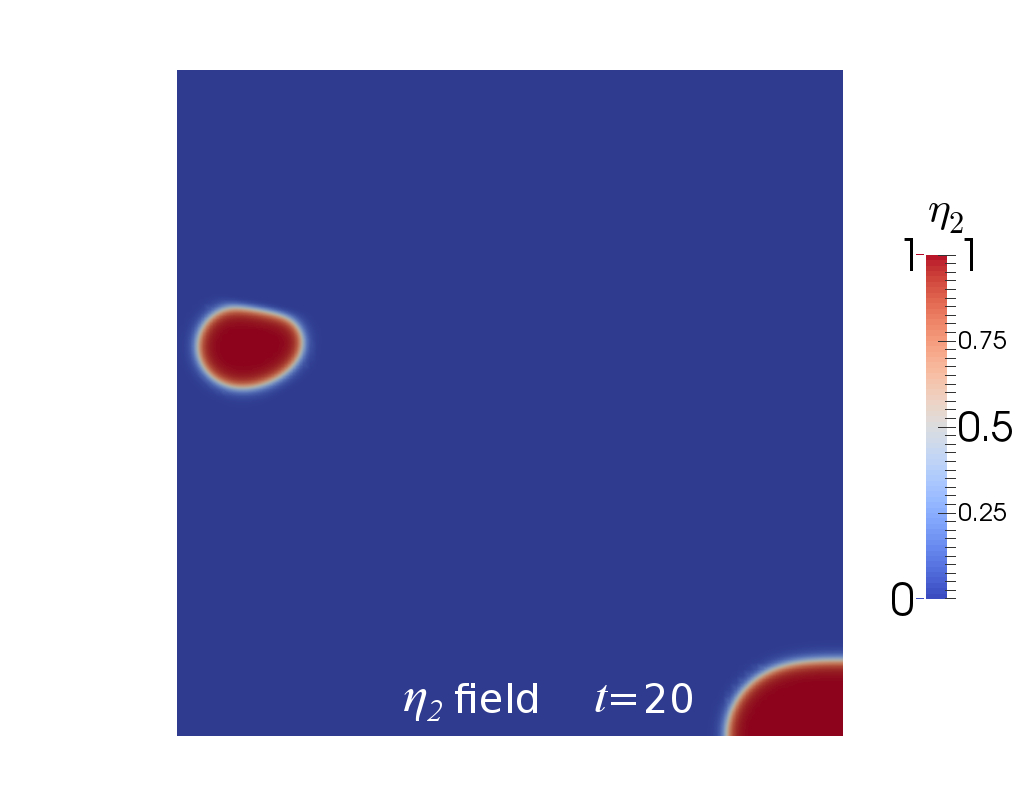}
\par\end{centering}

}\subfloat[]{\begin{centering}
\includegraphics[scale=0.19]{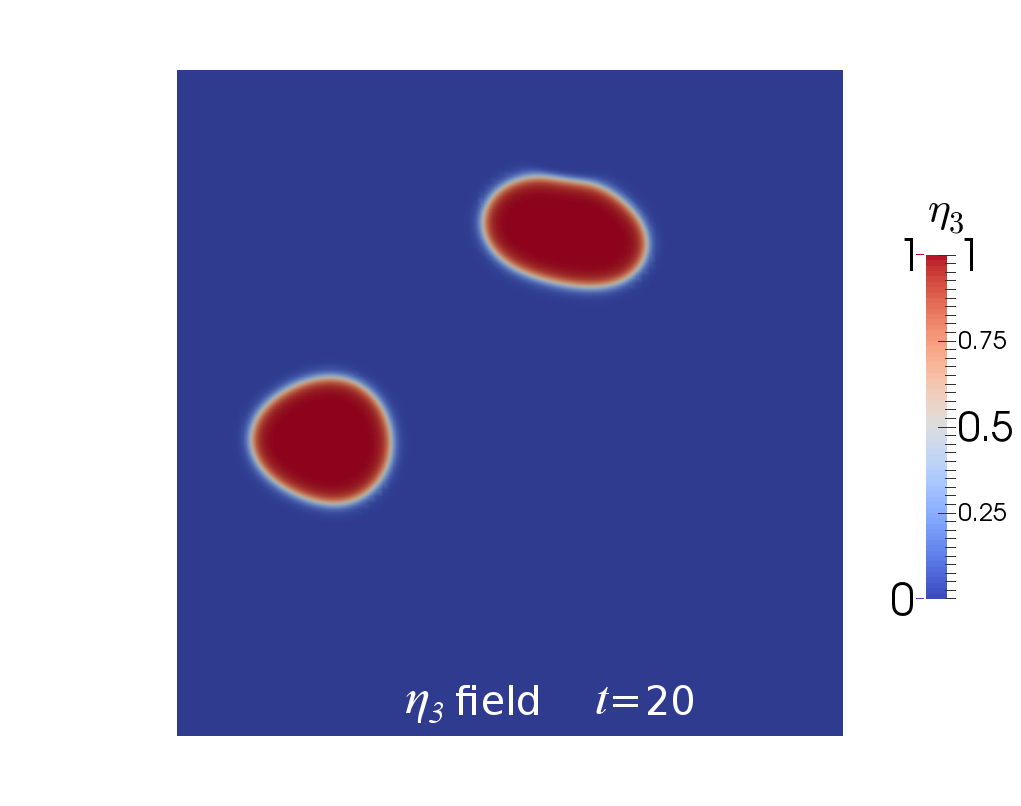}
\par\end{centering}

}\par\end{centering}

\begin{centering}
\subfloat[]{\begin{centering}
\includegraphics[scale=0.19]{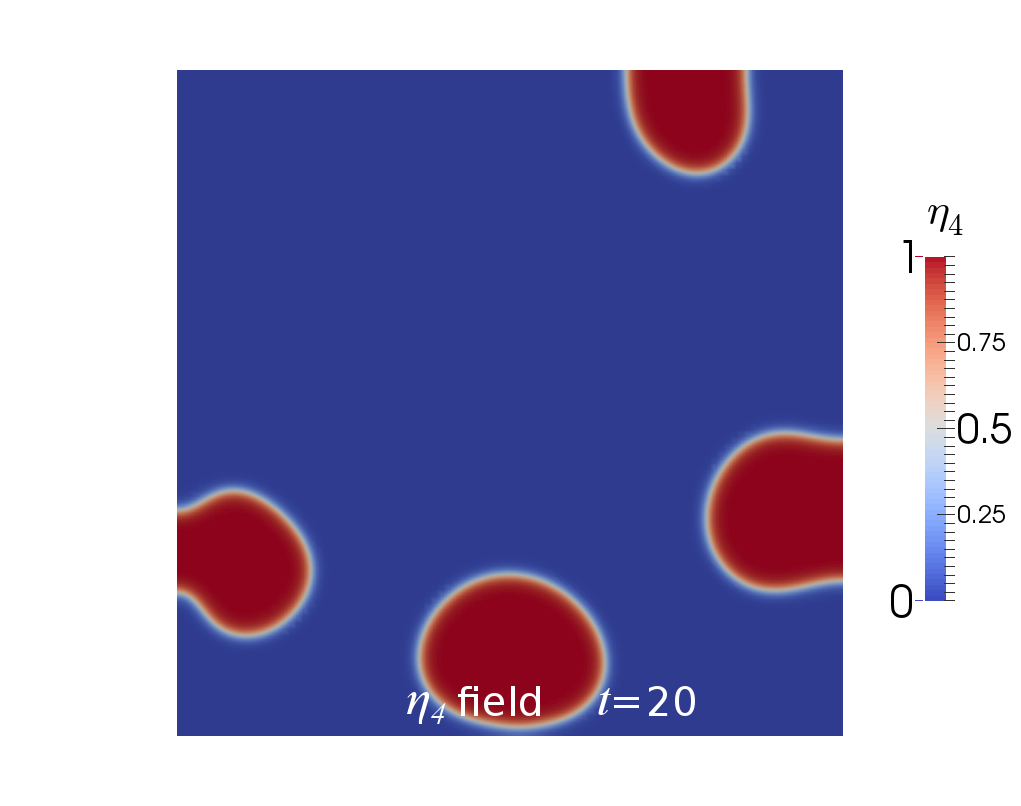}
\par\end{centering}
}
\par\end{centering}

\caption{The solute (a) and structural order parameter (b--e) fields at $t=20$ for the Ostwald ripening problem simulated with no-flux boundary conditions (no appreciable difference in results is observed between the conservative and aggressive time steppers). Second-phase particles of differing $\eta_i$ in contact with each other do not coalesce. \label{fig:p2_structure}}
\end{figure}

\begin{figure}
\begin{centering}
\subfloat[]{\begin{centering}
\includegraphics[scale=0.19]{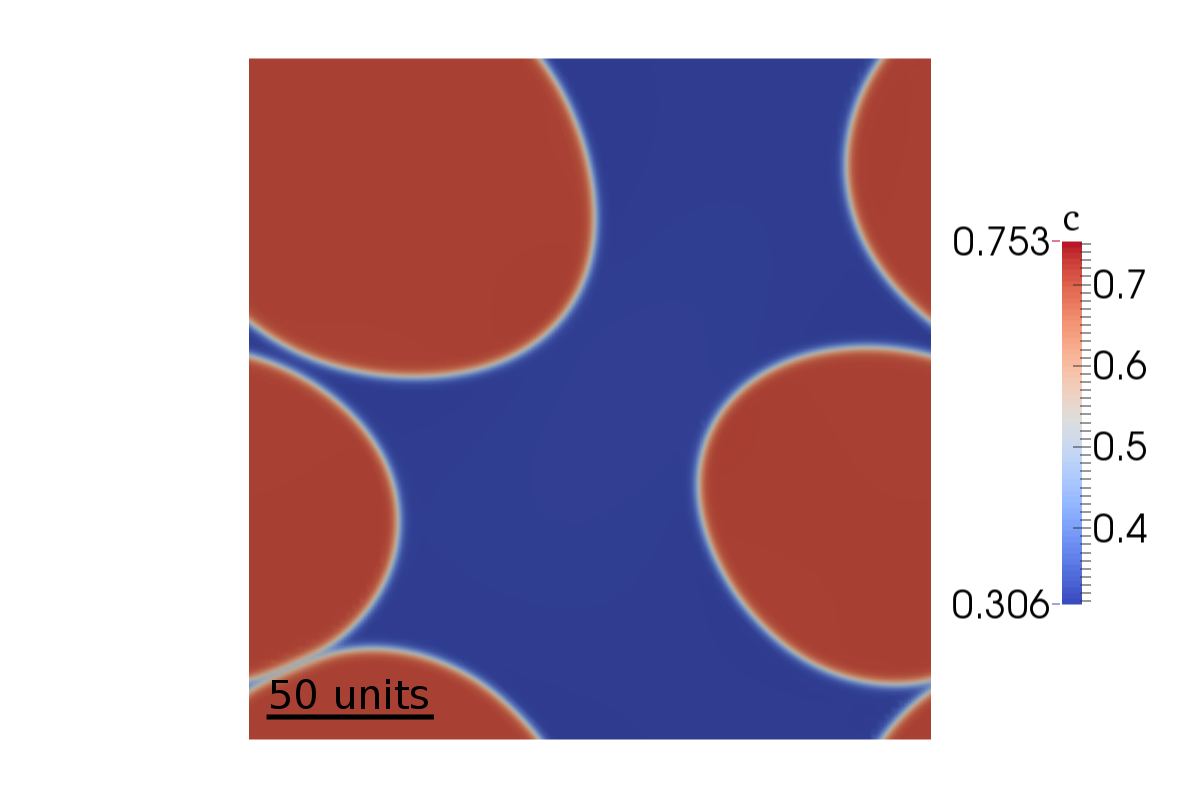}\label{fig:p2_IA}
\par\end{centering}

}\subfloat[]{\begin{centering}
\includegraphics[scale=0.25]{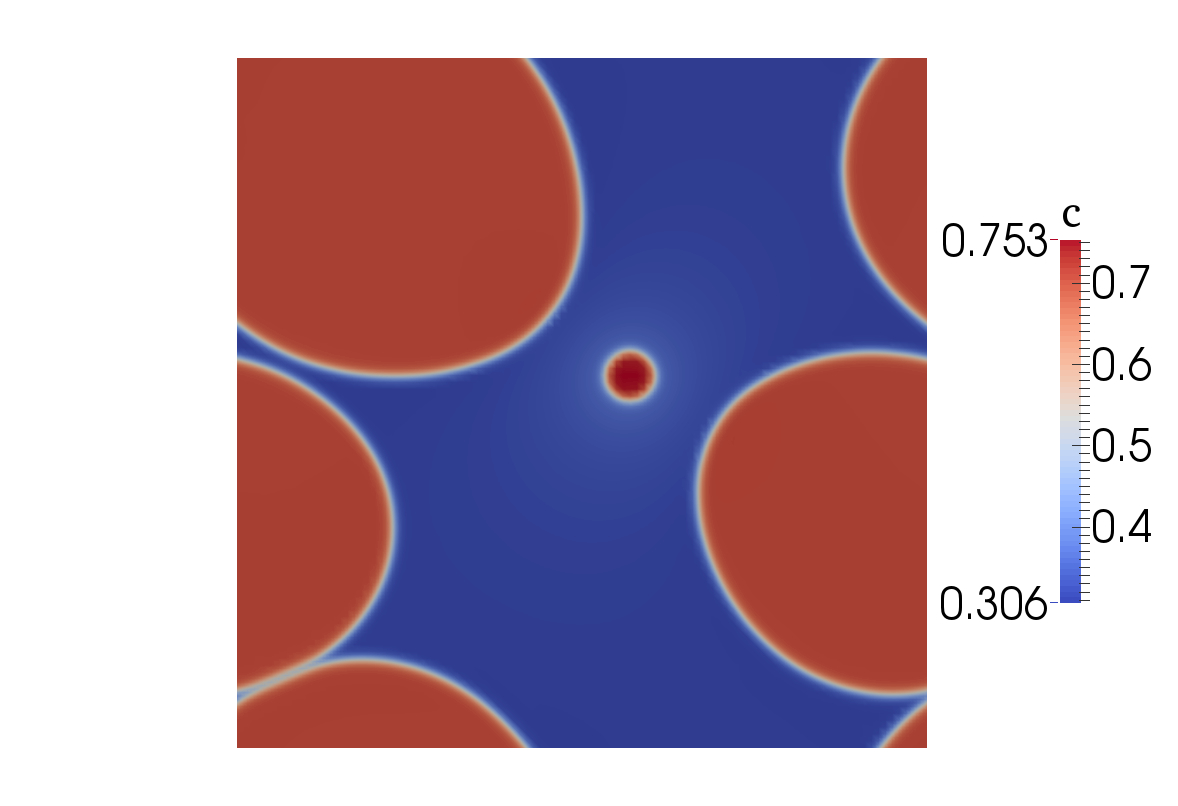}\label{fig:p2_STA}
\par\end{centering}

}
\par\end{centering}

\caption{A comparison of the composition fields of Ostwald ripening simulations performed with periodic boundary conditions illustrating the effect of the choice of time stepper on coarsening behavior. A shrinking particle has (a) completely dissolved by $t=4111$ when the simulation is performed with the conservative time stepper, while the particle has not yet completely dissolved by (b) $t=4131$ when the simulation is performed with the aggressive time stepper.   
\label{fig:coarsening_difference}}
\end{figure}

An example microstructure for the Ostwald ripening problem is shown in Fig.\ \ref{fig:p2_structure}, illustrating the solute and structural order parameter fields at $t=20$ for no-flux boundary conditions. The effect of the choice of time stepper is less evident in the free energy evolution (Fig.\ \ref{fig:p2_energy_evolution}), but in some cases, coarsening kinetics are impacted. Figure \ref{fig:coarsening_difference} shows parallel snapshots of the microstructure when the conservative and aggressive time steppers are applied to a simulation with periodic boundary conditions. In both simulations, a smaller particle in the center of the computational domain is shrinking; however, the particle has completely dissolved by $t=4111$ when the conservative time stepper is used (Fig.\ \ref{fig:p2_IA}), while it has not quite disappeared by $t=4131$ when the aggressive time stepper is used (Fig.\ \ref{fig:p2_STA}).  

As illustrated in Fig.\ \ref{fig:p2_energy_evolution}, the shrinkage of the central particle and the concomitant coarsening of the surrounding particles is slower when the simulation is performed using the aggressive time stepper, which is likely due to the fact that a particle shrinks faster as its radius decreases. The conservative time stepper naturally cuts the time step size as the rate of particle shrinkage increases, while the aggressive time stepper typically tends to increase (or at least maintain) the time step size until the solver is unable to converge to a solution. In multiple instances, particle dissolution is delayed, particularly in the later stages of the simulations.  This is likely due to the fact that at later stages of the simulation, there is a greater disparity in the radii of the shrinking and growing particles. In the case of late-stage particle shrinkage and dissolution, then, the aggressive time stepper may choose a time step size that is inappropriate for the physics of the system, increasing the error in the simulation. These results highlight the fact that adaptive time steppers must be carefully assessed and chosen to minimize their impact on the simulated microstructural evolution.

For these benchmark problems, we are interested in the microstructural evolution all the way to the lowest energy state, although for other problems, evolving to equilibrium or local energy minimum may be unrealistic or even uninteresting.  While the evolution of the total system energy provides important information to assess simulation results, we find that it is difficult to determine a proper simulation exit condition.  We originally tried a relative differential norm from one time step to the next with a tolerance of $5\times 10^{-8}$, but found that the simulations would sometimes exit significantly before equilibrium was reached.  Therefore, we ran the simulations without exit parameters and relied on human intervention. We found that once a simulation has visibly reached equilibrium (e.g., planar interfaces), the system free energy continues to decrease slowly, presumably due to equilibration of very small solute gradients. Eventually the free energy stops evolving to six or seven significant figures, at which point we chose to end the simulations.  In some simulations, however, the value of the total free energy fluctuates in the sixth or seventh significant digit, indicating that the solver has reached its limits in terms of numerical accuracy, and the simulations were again ended. 

\begin{figure}
\begin{centering}
\subfloat[]{\begin{centering}
\includegraphics[scale=0.7]{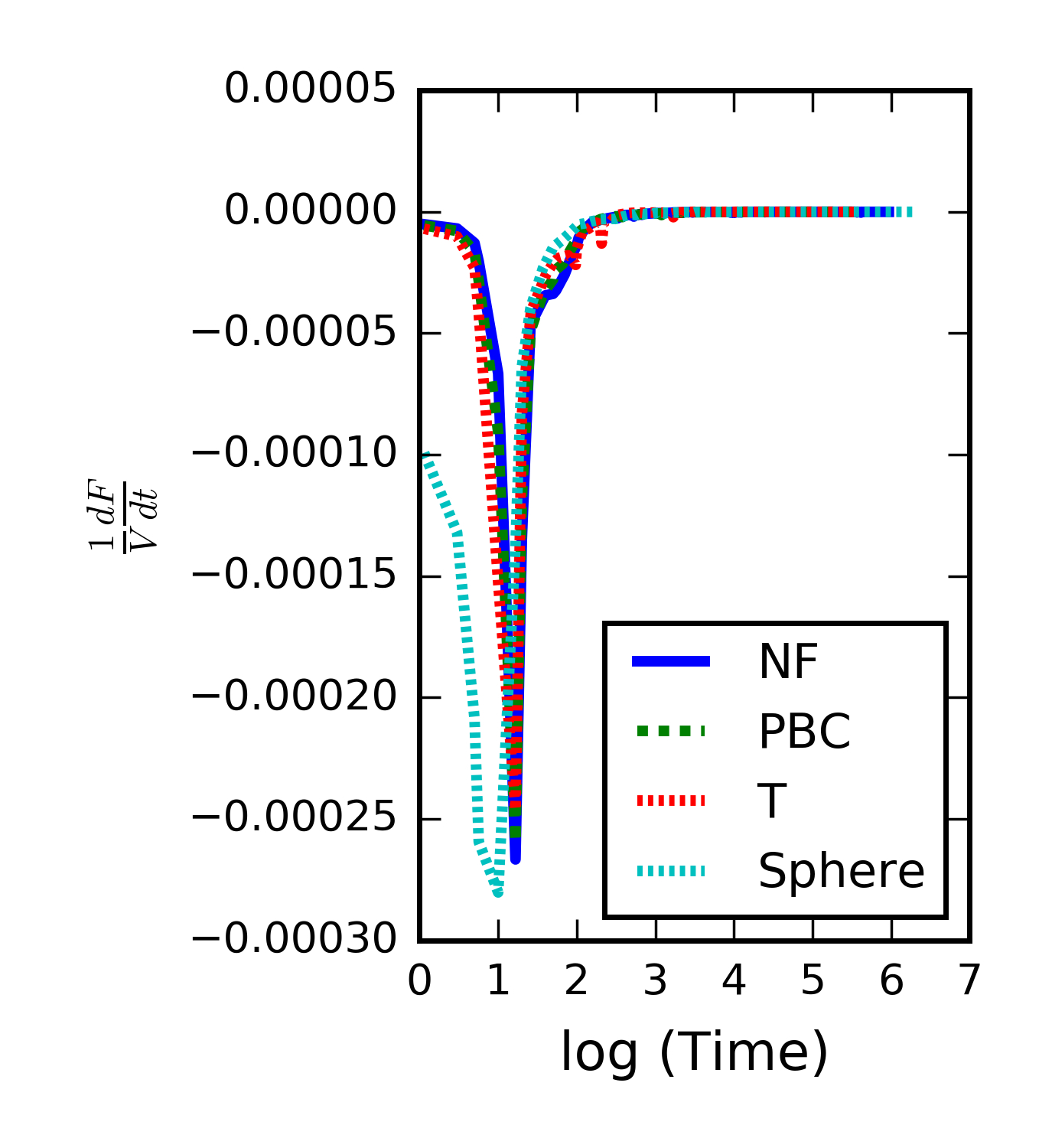}\label{fig:p1_dfdt_a}
\par\end{centering}

} \subfloat[]{\begin{centering}
\includegraphics[scale=0.7]{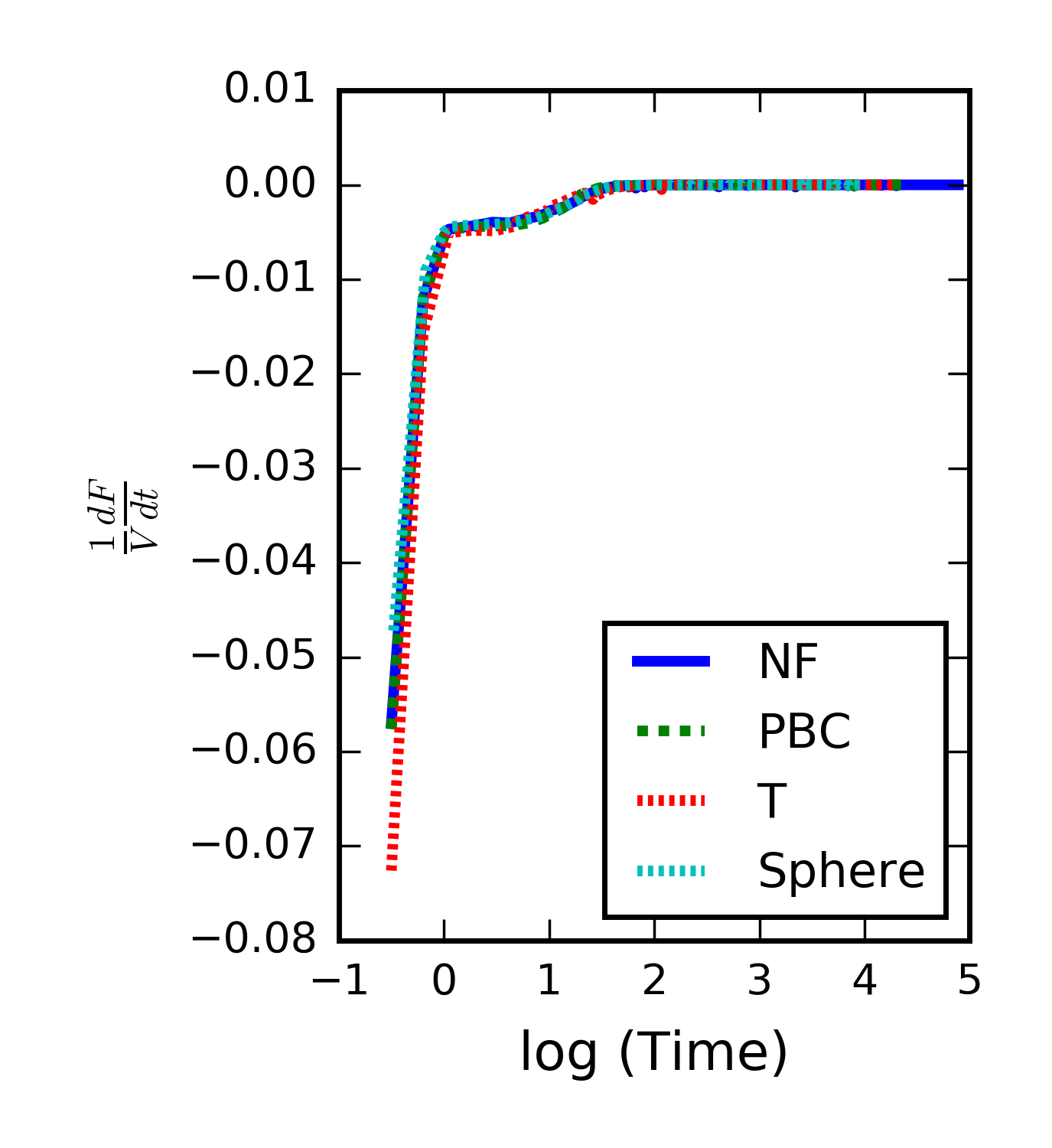}\label{fig:p1_dfdt_p2IA}
\par\end{centering}

} \subfloat[]{\begin{centering}
\includegraphics[scale=0.7]{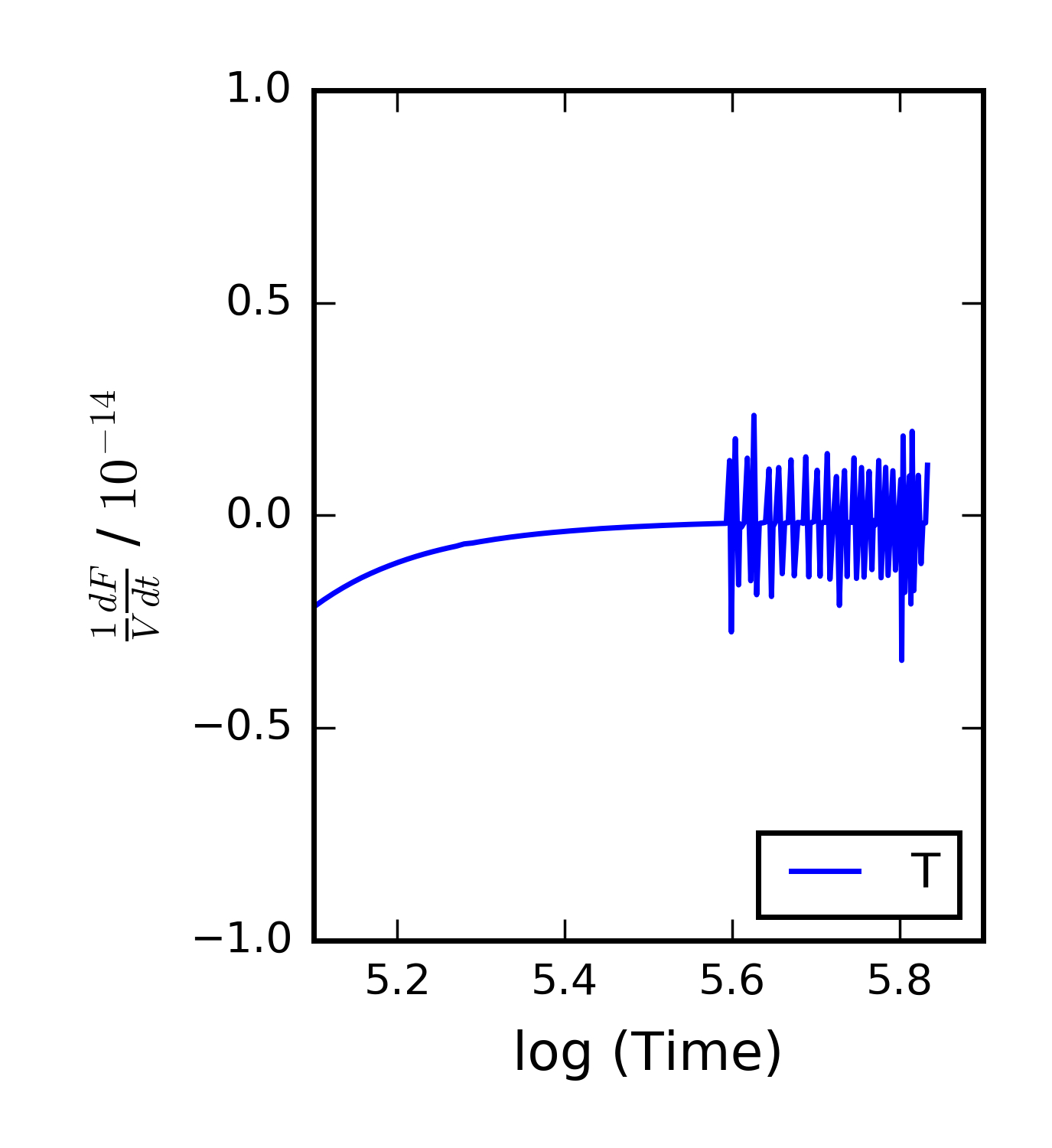}\label{fig:p1_dfdt_b}
\par\end{centering}

}\par\end{centering}

\caption{The calculated $\frac{1}{V}\frac{dF}{dt}$ for the benchmark problems simulated with the conservative time stepper, which  allows for a direct comparison of the evolution within the different computational domains: (a) spinodal decomposition, (b) Ostwald ripening.  (c) The calculated $\frac{1}{V}\frac{dF}{dt}$ for the T-shaped spinodal decomposition simulation as the simulation approaches equilibrium.  Note the numerical noise in the free energy of the system as it stops evolving. ``NF'' and ``PBC'' indicate square computational domains with no-flux boundary and periodic boundary conditions, respectively, and ``T'' and ``Sphere'' indicate the T-shaped and spherical surface computational domains.}
\label{fig:p1_dfdt}
\end{figure}

To determine a more useful criterion for ending these simulations, we calculate the rate of change of the volume-averaged free energy density, $\frac{1}{V}\frac{dF}{dt}$, an example of which is plotted in Fig.\ \ref{fig:p1_dfdt} for both benchmark problems simulated with the conservative time stepper. The rate of change allows for a more direct comparison of the evolution of the different within the different computational domains.  While the total free energies of the different simulations vary by several orders of magnitude because of their differing  computational domain size, $\frac{1}{V}\frac{dF}{dt}$ is similar, as shown in Fig.\ \ref{fig:p1_dfdt}.  The rate of change varies by about ten orders of magnitude throughout the course of the simulation, highlighting the need for accurate adaptive time stepping algorithms when studying long-term microstructural evolution. As shown for the T-shaped spinodal decomposition simulation in Fig.\ \ref{fig:p1_dfdt_b}, the numerical noise in the free energy (and thus $\frac{1}{V}\frac{dF}{dt}$) that can occur when the system stops evolving is evident.  As an example, we find that a value of $\frac{1}{V}\frac{dF}{dt}=1\times10^{-14}$ in dimensionless units generally appears to be sufficient for these benchmark problems to indicate that the final configuration has been achieved while avoiding the descent into numerical noise.  For other systems, the free energy and the rate of change of the average free energy density may need to be normalized to reasonable values for evaluation purposes.

The proposed benchmark problems presented in this paper model only a small subset of physics that have been incorporated into phase field formulations. Future benchmark problems should test additional key aspects, such as anisotropic linear elasticity with inhomogeneous moduli, anisotropic diffusivities and interfacial energies, solidification, and CALPHAD-based thermodynamics.  In addition, benchmark problems may benefit from being formulated with a parameter that controls the numerical difficulty of the problem, where possible.  The solvers and the interpretation of the problem statement may be verified for the ``easy'' problem, while the software may be stress-tested when the parameter value makes the problem ``difficult.'' Furthermore, perturbation studies, in which the initial conditions or problem parameters are slightly varied, may also be useful in determining how much simulation results differ as a result of numerical solvers versus any  inherent instability in the problem.  Finally, while the quantities of interest in each benchmark problem may vary (e.g., volume fraction of solidified material, polarization field in a ferroelectric material), we propose that the total free energy evolution is a standard, quantitative output that may be used for every problem. Additional relevant comparison metrics should be identified and utilized on a per-problem basis.  Community discussion and feedback is essential for the development of relevant, useful problem sets, and we urge individual researchers to contribute.

\section{Conclusion}

In this paper, we propose two benchmark problems for numerical implementations of phase field models that capture essential physical behavior present in a vast majority of models: solute diffusion and second-phase growth and coarsening. The model formulations are simplified to make the tests easier to implement within different software; however, the governing physics are captured. Furthermore, the initial conditions are formulated such that they are implementation-independent, yet still disordered, similar to initial conditions often used in the literature. Multiple computational domains and boundary conditions are given so that the numerical implementations may be challenged and to address future needs of phase field applications. We also discuss the need to produce tractable output, i.e., data formats that allow simulation results to be directly compared from different implementations, and identify the total free energy evolution as a metric that should be used for every problem. We demonstrate the utility of the benchmark problems by studying the effect of different time steppers on the microstructural evolution: small variations between the simulations at earlier times become amplified at later times. Given the deviation in our own results, we note that variations in results between different implementations does not necessarily imply an incorrect implementation. We also describe the use of the normalized rate of change of the total free energy to halt the simulations appropriately. Finally, the problems presented in this paper test only a small subset of the physics often incorporated into phase field models by design. Further benchmark problems are needed to model additional physics, such as linear elasticity, anisotropic diffusion and interfacial energies, solidification, and other phenomena. Ultimately, numerical benchmark problems should allow the validation of models with standard experimental data sets by ensuring that the differences in simulation results are not merely due to variations in numerical implementations.

For standard benchmark problems to become successful, the community must provide feedback about and input into the currently proposed problems and future problems. These standard benchmark problems are hosted on the NIST website, \url{https://pages.nist.gov/chimad-phase-field/}, along with the simulation data for the models presented here for download and comparison. The website will also serve as a repository for community-submitted results.  We encourage the community to contact the authors directly or via the website for additional discussion.

\section*{Acknowledgments}
The work by A.M.J., P.W.V., and O.G.H. was performed under financial assistance award 70NANB14H012 from U.S. Department of Commerce, National Institute of Standards and Technology as part of the Center for Hierarchical Material Design (CHiMaD). We gratefully acknowledge the computing resources provided on Blues and Fission, high-performance computing clusters operated by the Laboratory Computing Resource Center at Argonne National Laboratory and the High Performance Computing Center at Idaho  National Laboratory, respectively.  Finally, A.M.J. thanks J.R. Jokisaari for constructive writing feedback.


\bibliographystyle{model1-num-names}
\bibliography{hackathon.bib}


\end{document}